\documentclass[english,aps,pra,onecolumn,notitlepage,showpacs,superscriptaddress,nofootinbibfloatfix]{revtex4-1}
\usepackage[T1]{fontenc}
\usepackage[latin1]{inputenc}
\usepackage{graphicx}
\usepackage{bm}
\usepackage{amssymb}
\usepackage{color}
\usepackage[usenames,dvipsnames]{xcolor}
\usepackage{amsmath}
\usepackage{amstext}
\usepackage{latexsym}
\usepackage[colorlinks=true,citecolor=Cerulean,linkcolor=RubineRed,urlcolor=Cerulean]{hyperref}
\usepackage{lipsum}
\usepackage{enumitem}

\usepackage{mathptmx} 
\usepackage{verbatim}

\usepackage{amsfonts}
\usepackage{epsfig}
\usepackage{dsfont}
\usepackage{mathrsfs}
\usepackage{arydshln,leftidx,mathtools}

\begin{document}

\title{Quantum Supremacy of Many-Particle Thermal Machines}

\author{J. Jaramillo}
\affiliation{Department of Physics, University of Massachusetts, Boston, MA 02125, USA}
\affiliation{Department of Physics, National University of Singapore, Singapore 117551}
\author{M. Beau}
\affiliation{Department of Physics, University of Massachusetts, Boston, MA 02125, USA}
\affiliation{Dublin Institute for Advanced Studies, School of Theoretical Physics, Dublin 4, Ireland}
\author{A. del Campo}
\affiliation{Department of Physics, University of Massachusetts, Boston, MA 02125, USA}

\def\q{{\bf q}}

\def\G{\Gamma}
\def\L{\Lambda}
\def\la{\lambda}
\def\g{\gamma}
\def\al{\alpha}
\def\s{\sigma}
\def\e{\epsilon}
\def\k{\kappa}
\def\ve{\varepsilon}
\def\l{\left}
\def\r{\right}
\def\te{\mbox{e}}
\def\d{{\rm d}}
\def\t{{\rm t}}
\def\K{{\rm K}}
\def\N{{\rm N}}
\def\H{{\rm H}}
\def\la{\langle}
\def\ra{\rangle}
\def\om{\omega}
\def\Om{\Omega}
\def\vep{\varepsilon}
\def\wh{\widehat}
\def\tr{{\rm Tr}}
\def\da{\dagger}
\def\iz{\left}
\def\zi{\right}
\newcommand{\beq}{\begin{equation}}
\newcommand{\eeq}{\end{equation}}
\newcommand{\beqa}{\begin{eqnarray}}
\newcommand{\eeqa}{\end{eqnarray}}
\newcommand{\intf}{\int_{-\infty}^\infty}
\newcommand{\into}{\int_0^\infty}

\begin{abstract}
While the emergent field of quantum thermodynamics has the potential to impact energy science,
the performance of thermal machines  is often classical. We ask whether quantum effects can boost the performance of a thermal machine to reach quantum supremacy, i.e.,  surpassing both the efficiency and power achieved in classical thermodynamics. To this end, we introduce a nonadiabatic quantum heat engine operating an Otto cycle with a many-particle working medium, consisting of an interacting Bose gas confined in a time-dependent harmonic trap. 
It is shown that thanks to the interplay of nonadiabatic and many-particle quantum effects, this thermal machine can outperform an ensemble of single-particle heat engines with same resources, 
demonstrating the quantum supremacy of many-particle thermal machines. 
\end{abstract}

\maketitle

\section{Introduction}

\noindent The interplay of quantum technology and foundations of physics has turned quantum thermodynamics into a blooming field \cite{bookQThermo04,VA15}.
With the miniaturization to the nanoscale, a need  has emerged to understand and control the dynamics of thermal machines operating in the presence of both thermal and quantum fluctuations.
Quantum heat engines  (QHEs) constitute a prominent example, targeting the efficient conversion of heat into mechanical work.
The relevance of quantum thermal machines is further strengthened by their use to describe  both natural and artificial light harvesting systems 
 \cite{Scully10,Scully11,Dorfman13,Creatore13,Alicki15}.  
While quantum thermodynamics has the potential to revolutionize energy science, quantum thermal machines studied to date often exhibit a classical behavior.

Current efforts towards the realization of a tunable QHE in the laboratory use  a single-particle working medium, e.g., a confined ion in a modified Paul trap \cite{Abah12,Rossnagel14,Rossnagel15}. 
The optimization of this type  of single-particle QHE has received considerable attention \cite{YK06,Salamon09,Rezek09,Abah12,Deng13,Rossnagel14,Stefanatos14,Zhang14,delcampo14}. 
By contrast,  the performance of a QHE with a many-particle working medium remains essentially unexplored  \cite{Kim11,MD14,Campisi15,ZP15}. 
This is  however a timely question motivated by the prospects of scaling up QHE and related devices. 
In particular, an ion-trap  realization of a QHE constitutes a natural testbed to explore many-particle effects with well-established quantum technology, e.g., using an ion chain as a working medium.

In this article, we pose the question as to whether there exist scenarios in which the performance of a thermal machine exhibits {\it quantum supremacy}, i.e.,  a superior performance to that achievable in classical thermodynamics. 
To address this question we introduce a model of a many-particle QHE  and show that quantum supremacy can be achieved by  exploiting the interplay of nonadiabatic dynamics and many-particle quantum effects. 
In particular, we identify the conditions for which the use of a many-particle working medium leads to  a simultaneous enhancement of the efficiency and output-power of the many-particle QHE, 
surpassing the values for an ensemble of single-particle QHEs with same resources.

\begin{figure}
\begin{center}
\includegraphics[width=0.7\linewidth]{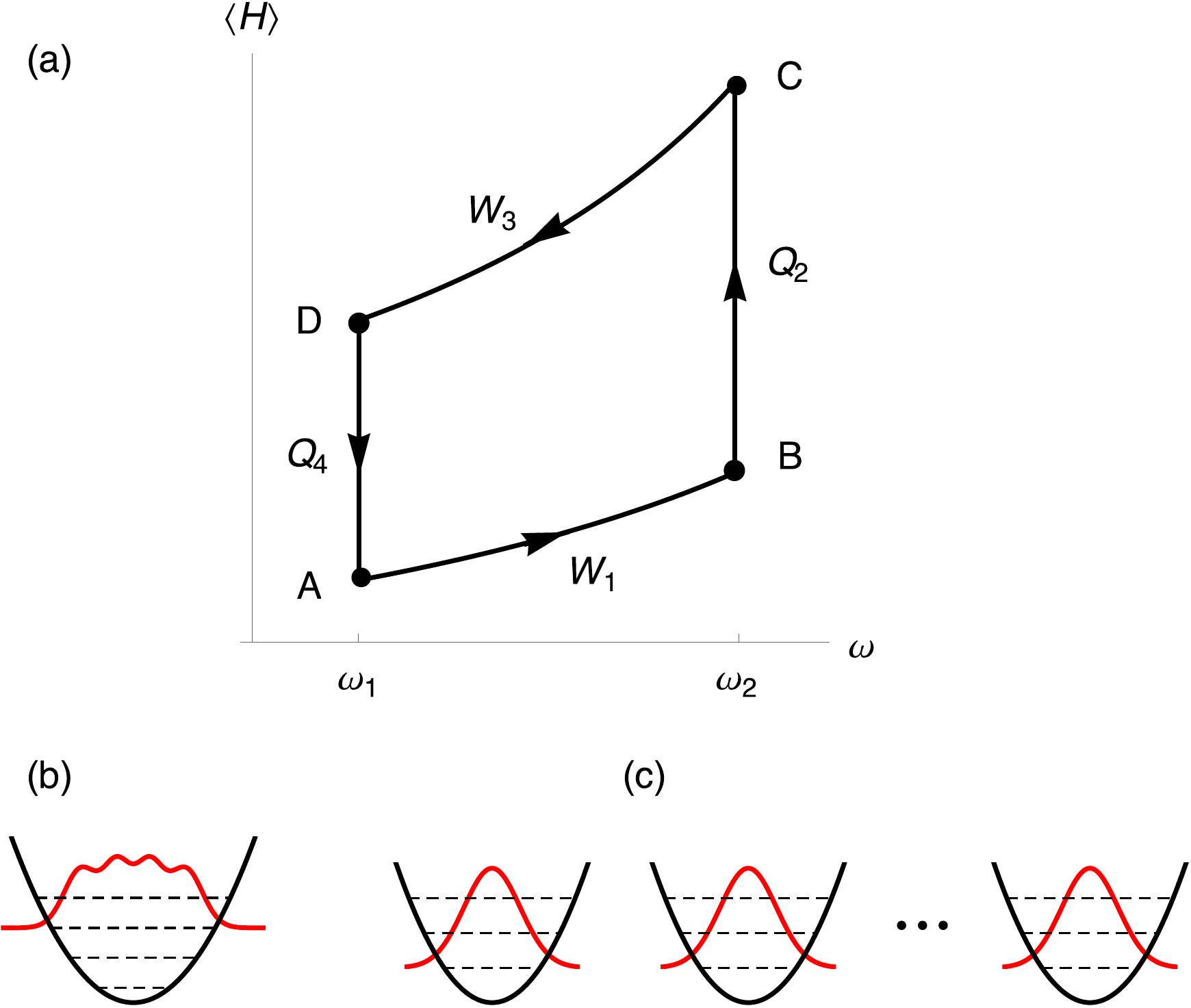}
\end{center}
\caption{\textbf{A quantum heat engine with an interacting Bose gas as a working medium.} (a) Quantum Otto cycle  for a working medium confined in a harmonic trap with frequency varying  between $\om_1$ and $\om_2$. States at A and C are thermal while those at B and D are generally nonequilibrium states.
(b) Schematic representation of the working medium in a many-particle heat engine compared to an ensemble of independent single-particle QHEs illustrated in (c). }
\label{figure1}
\end{figure}
%
%
%
The QHE we analyze operates an Otto cycle with a many-particle interacting quantum fluid in a time-dependent harmonic trap as a working medium that is alternately coupled to a hot and a cold reservoir, see Fig. \ref{figure1}. 
The Otto cycle is composed of four strokes shown in Fig. \ref{figure1}: 1) \textit{Adiabatic compression}: 
Starting from a thermal state $A$ with inverse temperature $\beta_{\rm c}$ and decoupled from any thermal reservoir, the working medium is driven by  increasing $\omega(t)$ from $\omega_1$ to $\omega_2$ until it reaches state $B$; 2) \textit{Hot isochore}: Keeping the trap frequency constant, the working medium is coupled to a hot reservoir at inverse temperature $\beta_{\rm h}$, relaxing to a thermal state $C$; 3) \textit{Adiabatic expansion}: The thermal state $C$ is decoupled from the hot reservoir and driven as $\omega(t)$  decreases  from  $\omega_2$ to $\omega_1$, reaching state $D$; 4) \textit{Cold isochore}:  Keeping  the  trap frequency $\om_1$ constant, the working medium is coupled to a cold reservoir at inverse temperature $\beta_{\rm c}$, relaxing to the initial thermal state $A$.

The efficiency of a heat engine is defined as the total output work per heat input
$
\eta=-\left[\langle W_1\rangle+\langle W_3\rangle\right]/\langle {\rm Q}_2\rangle.
$
 For the Otto cycle the input (output) work of the engine is given respectively by 
 $\langle W_{1(3)}\rangle=\langle H\rangle_{B(D)}-\langle H\rangle_{A(C)}$.
 Similarly, the heat flow in (out) reads 
$\langle {\rm Q}_{2(4)}\rangle=\langle H\rangle_{C(A)}-\langle H\rangle_{B(D)}$.\\

\section{Results}

\subsection{Nonadiabatic many-particle QHE}
We consider the working susbtance to be described by a quantum many-body Hamiltonian 
 \begin{equation}
 H=\sum_{i=1}^{\N}\left[-\frac{\hbar^2}{2m}\nabla^2+\frac{1}{2}m\omega(t)^2 {\bf r}_{i}^{2}\right]+
\sum_{i<j}V({\bf r}_i-{\bf r}_j),
 \label{Hsu11}
\end{equation}
 where $\N$ is the total number of particles and $\omega(t)$ is the trap frequency of the isotropic harmonic confinement. 
We assume $V({\bf r}/b)=b^2V({\bf r})$ 
so that the unitary dynamics generated by (\ref{Hsu11}) during the expansion and compression strokes exhibits scale-invariance \cite{Gambardella75}. The time-evolution of an eigenstate $\Psi$ at $t=0$ is \cite{Gritsev10,delcampo11b}
\beqa
\Psi({\bf r}_{1},\dots,{\bf r}_{\N},t)=\mathcal{N}e^{i\phi}\Psi\left(\frac{{\bf r}_{1}}{b},\dots,\frac{{\bf r}_{\N}}{b},t=0\right),
\eeqa
where the time-dependent phase  reads $\phi=\sum_im\dot{b}{\bf r}_{i}^{2}/(2\hbar b)$ and we have neglected 
a global dynamical phase which plays no role in our analysis. 
The normalization constant is $\mathcal{N}=b^{-\N d/2}$ for a system in $d$ spatial dimensions. 
The scaling factor $b=b(t)>0$ is the  solution of the Ermakov differential equation
$
\ddot{b}+\om(t)^2 b=\om_0^2/b^{3},
$
where we choose the constant $\om_0= \om(0)$ to be the  trap frequency at the beginning of the stroke. 
As shown in Appendix \ref{ExactManyBody1}, following a modulation of the trap frequency the mean nonadiabatic energy reads
$
\la H(t)\ra=Q^{\ast}(t)\la H(t)\ra_{\rm ad},
$
where 
\beqa
\label{qasteq}
Q^{\ast}(t)=b_{\rm ad}^2\left[\frac{1}{2b(t)^{2}}+\frac{\om(t)^2 b(t)^2}{2\om_0^2}+\frac{\dot{b}(t)^2}{2\om_0^2}\right],
\eeqa
is the nonadiabatic factor that accounts for the amount of energy excitations over the adiabatic dynamics,
  and $b_{\rm ad}=[\om_0/\om(t)]^{1/2}$ is the scaling factor in the adiabatic limit, obtained from the Ermakov equation by setting $\ddot{b}(t)\approx 0$. 
In this limit, $Q^{\ast}(t)$  reduces to unity and  the  mean energy is given by $\la H(t)\ra_{\rm ad}= \la H(0)\ra/b_{\rm ad}^2$. 
From the stationarity of the initial state it follows that  $b(0)=1$ and $\dot{b}(0)=0$.
Remarkably, $Q^{\ast}(t)$ coincides with the nonadiabatic factor introduced by Husimi for the single-particle time-dependent harmonic oscillator \cite{Husimi53}, as we show in  Appendix \ref{Husimi}.
As a result, at the end of the strokes decoupled from the heat reservoirs,
$
\langle H\rangle_B=Q^{\ast}_{AB}b_{\rm ad}^{-2}\langle H\rangle_{A}$ and $\langle H\rangle_D=Q_{CD}^{\ast}b_{\rm ad}^2\langle H\rangle_{C},
$
where $b_{\rm ad}=(\omega_1/\omega_2)^{1/2}$ and $Q_{AB(CD)}^*$ is the nonadiabatic factor for the compression (expansion) stroke.
It follows that the total input work per cycle $\langle W\rangle=\langle W_1\rangle+\langle W_3\rangle$ reads, 
\begin{eqnarray}\label{TotalWork}
\langle W\rangle&=&\left(Q_{AB}^{\ast}\frac{\omega_2}{\omega_1}-1\right)\langle H\rangle_A+\left(Q_{CD}^{\ast}\frac{\omega_1}{\omega_2}-1\right)\langle H\rangle_C.
\end{eqnarray}
From a practical point of view, the output power of the engine is more interesting as it accounts for the work  per cycle 
$
P\equiv -\langle W\rangle/(\tau_1+\tau_2+\tau_3+\tau_4)\ ,
$
where $\tau_i,\ i=1,\cdots,4$ corresponds to the duration of each stroke of the Otto cycle. 
Similarly, the efficiency of the many-particle QHE run in finite-time is found to be
\begin{eqnarray}\label{Efficiency}
\eta=1-\frac{\omega_1}{\omega_2}\left(\frac{Q_{CD}^{\ast}\langle H\rangle_C-\frac{\omega_2}{\omega_1}\langle H\rangle_A}{\langle H\rangle_C-Q_{AB}^{\ast}\frac{\omega_2}{\omega_1}\langle H\rangle_A}\right).
\end{eqnarray}
 The maximum efficiency is achieved under slow driving in the adiabatic limit
when the QHE operates at vanishing output power $P$ as a result of the requirement for a long cycle time $\tau=\sum_{i=1}^4\tau_i$. In this limit ($Q^{\ast}_{AB(CD)}\rightarrow 1$ or equivalently $|\dot{\om}/\om^2|\rightarrow 0$, see \cite{LR69}) equation (\ref{Efficiency}) reduces to the Otto efficiency $\eta_O=1-\omega_1/\omega_2$ which is shared as an upper bound by the single-particle quantum Otto cycle. 
By contrast, realistic engines operating in a finite time achieve a finite output power at the cost of introducing 
nonadiabatic energy excitations that represent  quantum friction. The engine efficiency is reduced as $Q_{AB(CD)}^{\ast}\geq 1$,
\begin{equation}\label{UBeta}
\eta\leq \eta_{{\rm nad},O}\equiv 1-Q_{CD}^{*}\frac{\omega_1}{\omega_2}.
\end{equation}
This bound is independent of the nature of the working medium (number of particles and inter-particle interactions) and is tighter than the Otto efficiency bound $\eta\leq\eta_O$. Equations \eqref{TotalWork}, \eqref{Efficiency}, and \eqref{UBeta} are the first results of this article giving explicit formulae for the total output work and efficiency, and showing the fundamental bound on the efficiency of a non-adiabatic QHE with the wide family of systems (\ref{Hsu11}) as a working medium \cite{Gritsev10,delcampo11b}.
We have recently discussed the optimization of a many-particle QHE using shortcut to adiabaticity \cite{BJdC16}.
In what follows we focus on enhancing the performance of the QHE without external control and derive the conditions  (number of particles, interaction strength, trap frequencies, as well as the temperatures of the hot and cold reservoirs) for quantum supremacy.  \\

\subsection{Quantum supremacy}
To explore the interplay between the nonadiabatic dynamics and  many-body effects, we consider a working medium consisting of $\N$  bosons confined in an harmonic trap and interacting through an inverse-square pairwise potential \cite{Calogero71,Sutherland71},
\begin{eqnarray}
 H=\sum_{i=1}^{\N}\left[-\frac{\hbar^2}{2m}\frac{\partial^2}{\partial z_i^2}+\frac{1}{2}m\omega(t)^2 z_{i}^{2}\right]+\frac{\hbar^2}{m}\sum_{i<j}\frac{\lambda(\lambda-1)}{(z_i-z_j)^2},
 \label{eq:csm}
\end{eqnarray}
where  $\lambda\geq 0$ is the coupling strength of the inter-particle interaction. 
This instance of (\ref{Hsu11}) is the Calogero-Sutherland gas and describes an ideal Bose gas for $\lambda=0$ as well as  hard-core bosons (Tonks-Girardeau gas) for $\lambda=1$ \cite{Girardeau60,GWT01}. The thermodynamics of the latter is identical to that of polarized fermions. For other values of $\lambda$, the Hamiltonian \eqref{eq:csm} describes hard-core bosons with inverse-square interactions and can be reinterpreted as an ideal gas of particles obeying generalized-exclusion statistics \cite{Haldane91,Wu94,MS94}.
While the eigenstates are a paradigm of strong correlations with Bijl-Jastrow form, the spectrum is purely linear \cite{Calogero71,Kawakami93}, 
$
E(\{n_j\})=(\hbar\omega\N/2)[1+ \lambda(\N-1)] +\sum_{j=1}^{\infty} j \hbar \omega n_{j},
$
where  $n_j$ is  the occupation number in the  $j$-th mode of energy $j\hbar\om$, satisfying the normalization $\sum_{j=1}^{\infty}n_j=\N$.  
To determine the output power and efficiency of the QHE, we first note that the partition function is given by $Z_\N^{(\lambda)}=e^{-\beta\lambda\frac{\hbar\omega}{2}\N(\N-1)}Z_\N^{(0)}$ where the partition function for an ideal Bose gas $Z_\N^{(0)}=e^{-\beta\hbar\omega/2}\prod_{k=1}^{\N}(1-e^{-\beta k\hbar\omega})^{-1}$ can be computed via recurrence relations \cite{MF03}, see Appendix \ref{GeneralResults1} for further details.
Knowledge of the partition function allows one to compute the equilibrium mean thermal energy via the identity $\langle H\rangle=-\partial_{\beta} {\rm ln} Z^{(\lambda)}_{\N}$, i.e.,
$
\langle H\rangle=\frac{\hbar\omega}{2} \N[1 +\lambda(\N-1)]+\hbar\omega\sum_{k=1}^{\N}  k(e^{\beta k\hbar\omega}-1)^{-1}.
$
In the adiabatic limit, $Q^{\ast}(t)=1$, both the output power per cycle and the efficiency become independent of the interaction strength $\lambda$, see Appendix \ref{GeneralResults2}. By contrast, for an arbitrary nonadiabatic driving protocol with $Q^{\ast}(t)>1$, they both decrease monotonically as a function of $\lambda$. It follows that the output power is optimal for an ideal Bose gas. 
As a paradigm for nonadiabatic effects, we shall consider in the following a sudden-quench driving of the trap frequency between $\om_1$ and $\om_2$. In this case, the nonadiabatic factor takes the form $Q^{\ast}_{AB(CD)}=(\omega_1^2+\omega_2^2)/(2\omega_1\omega_2)$ that is symmetric with respect to the exchange  $\om_1\leftrightarrow\om_2$.
As a result, the efficiency (\ref{Efficiency}) of a realistic QHE with a short time per cycle has as fundamental upper limit \eqref{UBeta}, $\eta_{\text{sq}}\leq \eta_{{\rm nad},O}=(1-(\omega_1/\omega_2)^2)/2$, i.e., not higher than $50\%$. Despite this limit,  we show next that nonadiabatic quantum effects can enhance the efficiency of a multi-particle QHE over the single-particle counterpart.

\begin{figure*}[t]
\begin{center}
\includegraphics[width=0.4\linewidth]{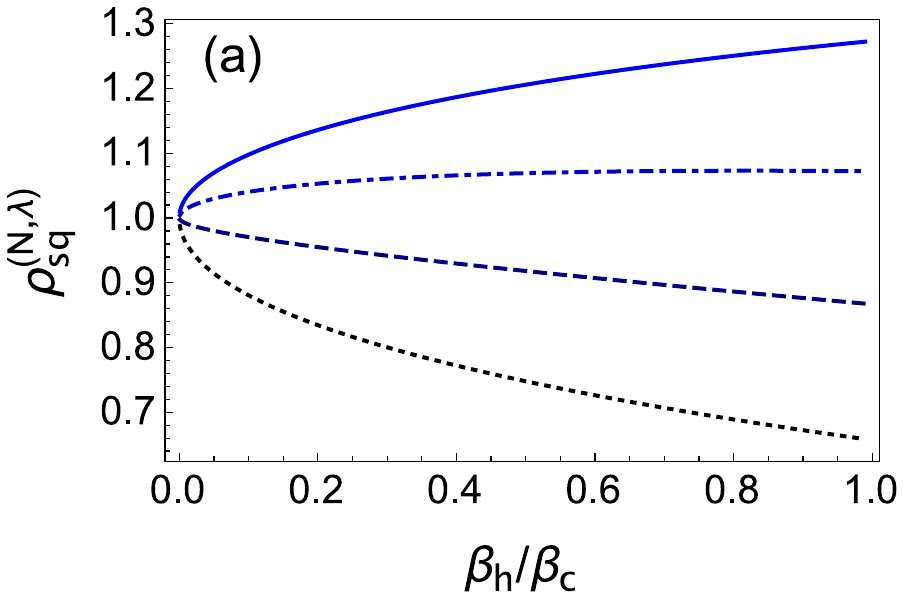}\ \ \ \includegraphics[width=0.4\linewidth]{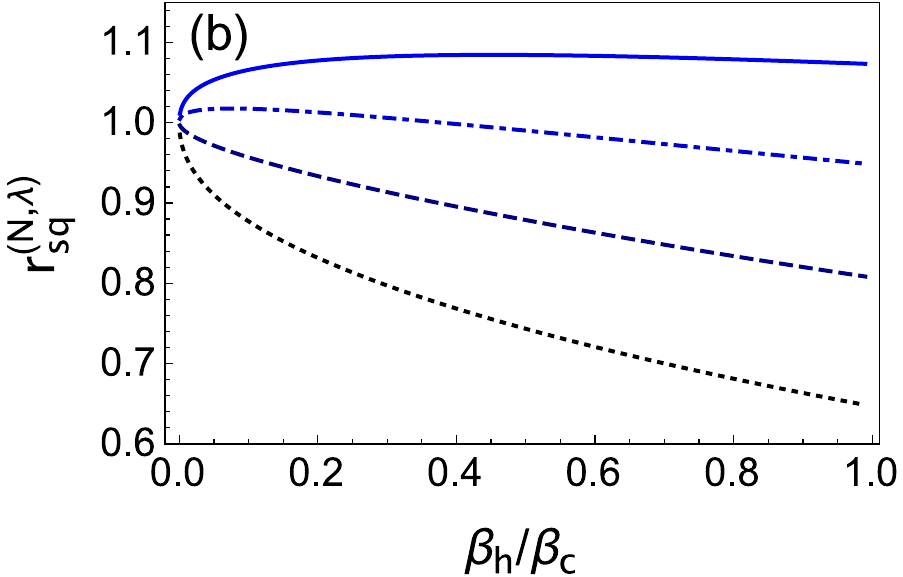}
\caption{\textbf{Effect of the interactions on the performance of a nonadiabatic many-particle QHE compared to an ensemble of $\N$ single-particle QHEs.} 
(a) Efficiency ratio at optimal output power, see equation \eqref{ratio}, as a function of $\beta_{\rm h}/\beta_{\rm c}$ with $\beta_{\rm c}=0.01/(\hbar\om_1)$ and $\N=200$ ($\sigma_c=2$) for $\lambda=0,0.2,0.5,1$ from bottom to top. A sudden-quench protocol is considered.
(b) Corresponding output power ratio under  the same conditions.
We observe that the ideal Bose gas $\lambda=0$ as well as for the weakly interacting Bose gas $\lambda=0.2$ one can boost the performance of a many-particle QHE for $\N$ large while it is better to engineer an ensemble of single-particle QHEs when the interaction strength $\lambda \geq 1/2$. }
\label{figure2}
\end{center}
\end{figure*}

\begin{figure*}[t]
\begin{center}
\includegraphics[width=0.4\linewidth]{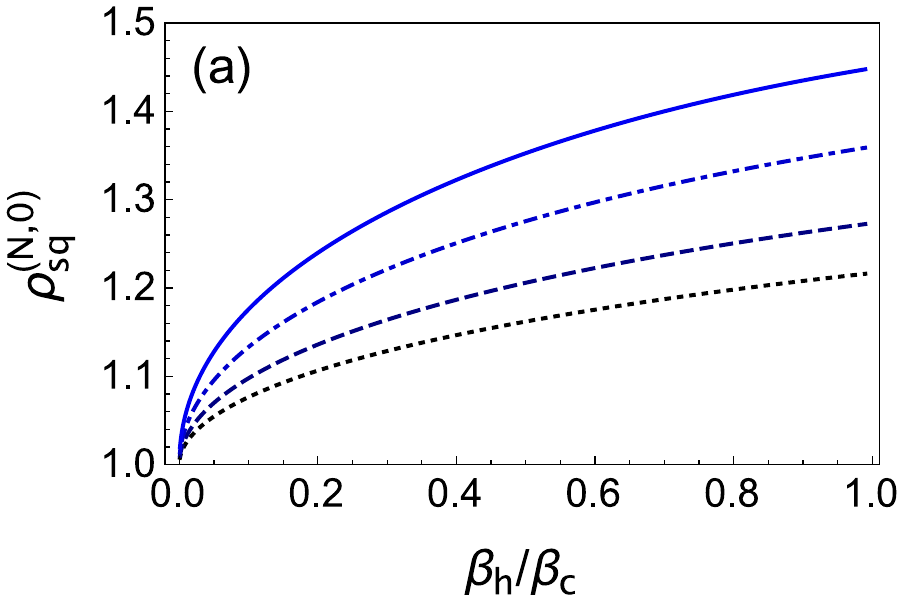}\ \ \ \includegraphics[width=0.4\linewidth]{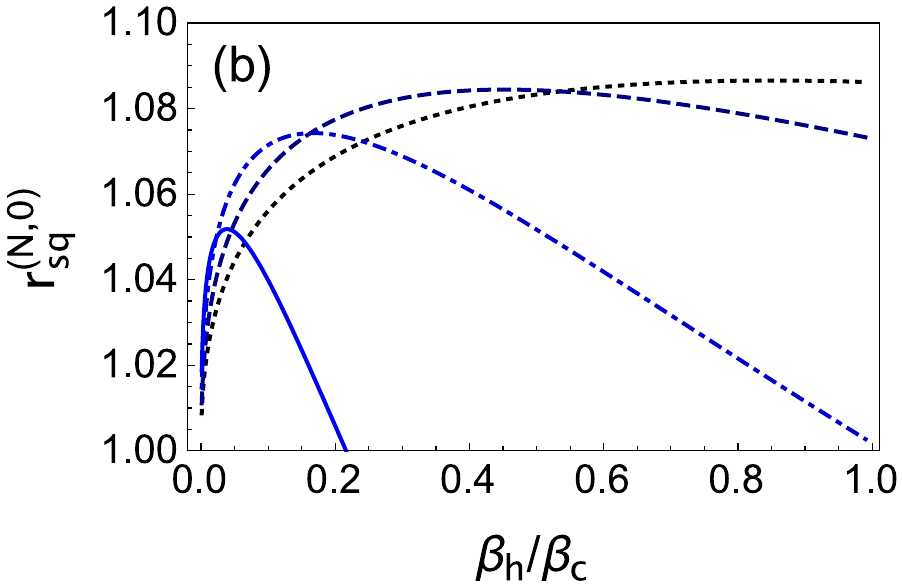}
\includegraphics[width=0.4\linewidth]{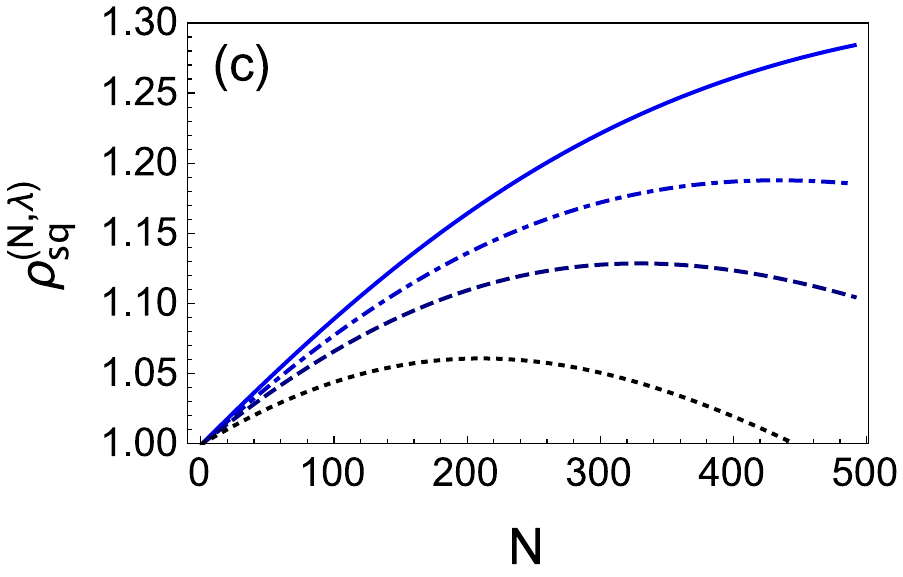}\ \ \ \includegraphics[width=0.4\linewidth]{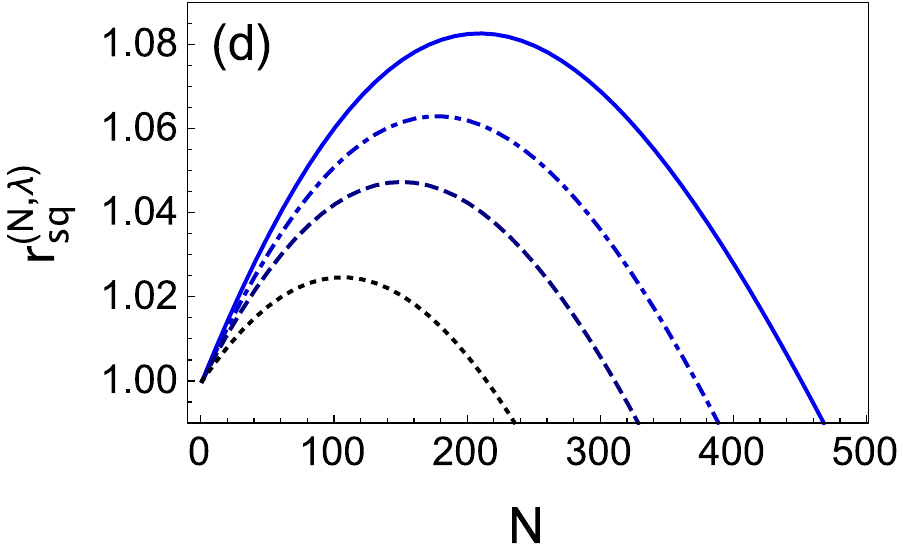}
\caption{\textbf{Quantum supremacy of a nonadiabatic many-particle QHE compared to an ensemble of $\N$ single-particle QHEs.} 
Efficiency ratio (left) and output power ratio (right)  at optimal output power, see equation \eqref{ratio}, as a function of $\beta_{\rm h}/\beta_{\rm c}$ (respectively $\N$) with $\beta_{\rm c}=0.01/(\hbar\om_1)$: 
(a)-(b) Robustness of quantum supremacy for different values of  the particle number of an ideal Bose gas ($\lambda=0$) with $\N=150$ (dotted curve), $200$ (dashed curve), $300$ (dashed-dotted curve), $500$ (continuous curve), corresponding to $\sigma_c=1.5,2,3,5$ respectively. 
(c)-(d) Dependence on the interaction strength $\lambda=0,0.05,0.1,0.2$ from top to bottom, for  a fixed ratio of temperatures $\beta_{\rm h}/\beta_{\rm c}=0.3$. A sudden-quench driving is considered in all cases.}
\label{figure3}
\end{center}
\end{figure*}

We determine the conditions for optimal output power $P^{(\N,\lambda)}_{\rm sq}$ by optimizing the ratio $\om_1/\om_2$ for fixed $\om_1$, temperatures $\beta_{\rm c},\ \beta_{\rm h}$, number of particles $\N$, and interaction strength $\lambda$. We denote the efficiency at optimal output power by $\eta^{(\N,\lambda)}_{\rm sq}$.
To characterize many-particle effects, we introduce the following ratios
\begin{gather}\label{ratio}
r_{\rm sq}^{(\N,\lambda)}\equiv \frac{P^{(\N,\lambda)}_{\rm sq}}{\N\times P^{(1,\lambda)}_{\rm sq}},\ \ 
\rho_{\rm sq}^{(\N,\lambda)}\equiv \frac{\eta^{(\N,\lambda)}_{\rm sq}}{\eta^{(1,\lambda)}_{\rm sq}},
\end{gather}
where the first (second) ratio compares the optimal output power (efficiency at optimal output power) of a $\N$-particle QHE with that of $\N$ single-particle QHEs. Their dependence on the interaction strength is displayed in Fig. \ref{figure2}, where it is shown that they can be greater than 1 for $\lambda < 1/2$ and optimal for $\lambda=0$.
In our following analysis, we will show that for some temperature regimes, the single-particle QHE exhibits classical performance up to first order corrections, while the many-particle QHE can retain a leading quantum term of the same magnitude as the single-particle classical one. In this regime, the two ratios defined in equation \eqref{ratio} compare the quantum many-particle performance with that of an ensemble of classical single-particle QHEs. 
We consider that a QHE exhibits  quantum supremacy  when both ratios given in equation \eqref{ratio} are greater than one, see Fig. \ref{figure3}. 
We have verified that the conditions for quantum supremacy do not substantially depend on the definition of (\ref{ratio}). In particular,  a similar definition in which the efficiency and power of both single- and many-particle QHEs are evaluated at the same frequency ratio yields essentially the same conditions for quantum supremacy, see Appendix \ref{AppendixQS}.
To analyze the optimal output power we resort to the numerically-exact optimization shown in Figs. \ref{figure2} and \ref{figure3} where the ratios in equation \eqref{ratio} are plotted for different values of $\N$ and $\lambda$. The performance of a many-particle QHE is shown to depend strongly on the regime of temperature, particle number, and interaction strength.

In the high-temperature limit of the hot reservoir  $\sigma_{\rm h}\equiv\N\hbar\beta_{\rm h}\omega_2\ll 1$ 
and of the cold reservoir $\sigma_{\rm c}\equiv \N\hbar\beta_{\rm c}\omega_1\ll 1$, we find that the optimal frequency is given by $(\omega_1/\omega_2)_{\text{sq}}\approx a^{1/4}[1+\frac{1}{8}g_{\N}(\lambda-\frac{1}{2})\sigma_{\rm c}]$ where $g_{\N}=(\N-1)/\N$ and $a\equiv \beta_{\rm h}/\beta_{\rm c}$ denotes the ratio of temperatures (see Appendix \ref{SuddenQuench}  for details), whence it follows that 
\begin{equation}
\eta_{\text{sq}}^{(\N,\lambda)}=\frac{1-\sqrt{a}}{2+\sqrt{a}}+\frac{(3-2a^{\frac{3}{4}})\sqrt{a}}{4(2+\sqrt{a})^2}g_\N\left(\frac{1}{2}-\lambda\right)\sigma_{\rm c} \label{etaSQ:sigmasmall}.
\end{equation}
As a result, we recognize the leading term in $\eta_{\text{sq}}^{(\N,\lambda)}$, independent of $\sigma_{\rm c}$, as the Rezek-Kosloff efficiency shared by single-particle heat engines under sudden driving \cite{YK06}. 
The leading quantum $\sigma_{\rm c}$-contribution  to the efficiency at optimal work increases with the number of particles $\N$  for  $\lambda < 1/2$. On the contrary, if $\lambda > 1/2$  quantum effects lower the efficiency and vanish for $\lambda=1/2$ or $\N=1$. A similar result can be found for the optimal output power. 
We note that for an adiabatic protocol quantum corrections of this order vanish and the engine operates at the Curzon-Ahlborn efficiency \cite{CA75}, see details in Appendix \ref{Adiabatic}. 
Equation \eqref{etaSQ:sigmasmall} suggests that for a QHE with a small particle number, quantum corrections can lead to quantum supremacy. Yet, $r_{\rm sq}^{(\N,\lambda)}$ and $\rho_{\rm sq}^{(\N,\lambda)}$ vary as $1+O(\sigma_{\rm c})$ where the corrections depend strongly on $\lambda$ and $\N$. Numerically, one can see that quantum deviations predicted in equation \eqref{etaSQ:sigmasmall} amount to a few percents with respect to the single-particle case. 

In the opposite limit of very low temperature $\sigma_{\rm c}\gg \N$ of the cold reservoir (keeping at high temperature the hot reservoir $\sigma_{\rm h}\ll \N$), the thermal energy of the equilibrium state $A$ does not contribute to the total work. One has $(\omega_1/\omega_2)_{\text{sq}}\approx(\kappa_{\N,\lambda}^2\hbar\beta_{\rm h}\omega_1)^{1/4}$ (see Appendix \ref{SuddenQuench} for details),  
where $\kappa_{\N,\lambda}=\sqrt{(1+(\N-1)\lambda)}$. The efficiency at optimal output power reads
\begin{equation}
\eta_{\text{sq}}^{(\N,\lambda)}=\frac{1-\kappa_{\N,\lambda}\sqrt{\hbar\beta_{\rm h}\omega_1/2}}{2+\kappa_{\N,\lambda}\sqrt{\hbar\beta_{\rm h}\omega_1/2}}.\label{etaSQ:LT}
\end{equation}
For $\N=1$, equation (\ref{etaSQ:LT}) reduces to the efficiency reported for a single-particle quantum Otto cycle \cite{Abah12}. In this regime, the efficiency at optimal output power decreases drastically with respect to the high temperature regime, see equation \eqref{etaSQ:sigmasmall}, 
because $\hbar\beta_{\rm h}\omega_1\gg \beta_{\rm h}/\beta_{\rm c}$ by assumption. Therefore, it is not worth running the QHE in the very low temperature limit of the cold reservoir because the working substance is effectively ``frozen'' in the ground state, reducing the output work per cycle. Numerical simulations show that a similar conclusion holds in the regime $\sigma_{\rm c}\sim \N$ and for $\sigma_{\rm c},\sigma_{\rm h}\geq \N$.

We have seen that estimate \eqref{etaSQ:sigmasmall} is valid for small number of particles (e.g., $\N\sim 1-30$ for $\sigma_{\rm c}/\N=0.01$) and  values of the ratio $\beta_{\rm h}/\beta_{\rm c}$ consistent with the assumption $\sigma_{\rm h}\ll 1$. In this regime, the enhancement of the efficiency is limited to few percents of the single-particle case. In addition, for very low temperatures, where one could expect significant quantum corrections, the efficiency decreases drastically. Yet, many-particle quantum thermodynamics  exhibits a novel, intermediate regime accessible for a large number of particle and characterized by $1\leq \sigma_{\rm c}\ll \N$. This is a low temperature regime but admitting enough thermal excitations to prevent the working medium from being  effectively ``frozen'' in a single many-particle eigenstate (the ground state). In this regime, the energy $\langle H\rangle_{A(C)}\approx (\N/\beta_{\rm c(h)})\mu_\lambda(\sigma_{\rm c(h)})$ where $\mu_\lambda(\sigma)=\sigma^{-1}\int_{0}^{\sigma}s(e^s-1)^{-1}ds+\lambda\sigma/2$, $\sigma_{\rm c(h)}=\N\hbar\beta_{\rm c(h)}\omega_{1(2)}$, and $\mu_{\lambda}(\sigma_{\rm c(h)})$ represents the relative deviation to the classical value of the mean energy $\N/\beta_{\rm c(h)}$ for an equilibrium state in the canonical ensemble, see Appendix \ref{GeneralResults1}. Using \eqref{TotalWork} for a sudden quench driving we explore numerically the optimization of the output power $P=-\langle W\rangle/\tau$. It is clear that for large particle number, the efficiency at optimal output power deviates from the Rezek-Kosloff efficiency given by the leading term in equation \eqref{etaSQ:sigmasmall}. In this scenario, nonadiabatic effects during sudden-quench driving can be used to achieve quantum supremacy. In particular, they lead to a many-particle enhancement of the efficiency at optimal output power by up to $\sim$50$\%$ of the single-particle value (for $\sigma_{\rm c}\geq 10$, typically), see Appendix \ref{Multipart} for a derivation of this upper bound. In Fig. \ref{figure3} we observe that for $\lambda=0$ (free bosons) and for $\N\leq 300$, the optimal output power (and efficiency at optimal output power) of a many-particle QHE surpasses the value of $\N$ independent single-particle QHEs. For $\N\geq 300$ the relative efficiency at optimal output power increases with $\N$ while the many-particle optimal output power can be improved (with respect to $\N$ single-particle QHEs) for $\beta_{\rm h}/\beta_{\rm c}$ less than a critical value ($\sim 0.2$ for $\N=500$). For a weakly interacting Bose gas (typically $\lambda\leq 0.2$) the optimal output power and efficiency are boosted for a range of $\N$  below a critical value depending on the interaction strength $\lambda$ and also on temperatures $\beta_{\rm h}$ and $\beta_{\rm c}$. \\

\section{Discussion}

\noindent In this article, we have shown that quantum thermal machines  can exhibit quantum supremacy, i.e., a superior performance to that allowed in classical thermodynamics.
To this end, we have introduced a many-particle quantum heat engine with an interacting Bose gas confined in a time-dependent harmonic trap as a working medium. 
Our analysis shows that  quantum supremacy results from the interplay between many-particle quantum effects and nonadiabtic dynamics that boost the finite-time efficiency and output power of a many-particle QHE, 
surpassing that of an ensemble of single-particle heat engines, matched by the predictions of classical thermodynamics. Our results should motivate future experimental research in scaling up nanoscale thermal machines whereby nonadiabatic many-particle quantum effects are exploited to achieve quantum supremacy.\\

\noindent{\bf Acknowlegments} 
\noindent It is a great pleasure to thank  A. Chenu and B. Sundaram for illuminating discussions.
Funding support from UMass Boston (project P20150000029279) and ESF (POLATOM-5052) is further acknowledged.\\

\appendix

\section{Exact many-body dynamics and the  $SU(1,1)$ dynamical symmetry group}\label{ExactManyBody}

\subsection{Dynamical scaling for the nonadiabatic mean energy}\label{ExactManyBody1}

We consider a confined quantum fluid as a working medium with Hamiltonian
\begin{eqnarray}
 H=\sum_{i=1}^{\N}\left[-\frac{\hbar^2}{2m}\frac{\partial^2}{\partial z_i^2}+\frac{1}{2}m\omega(t)^2 z_{i}^{2}\right]+\sum_{i<j}V(z_i-z_j)\ ,
 \label{hscale}
\end{eqnarray}
where the inter-particle interactions are described by a potential that exhibits the  scaling property $V(\lambda z)=\lambda^{-2}V(z)$. The dynamical group for the Hamiltonian \eqref{hscale} is $SU(1,1)$ \cite{Gambardella75}. For simplicity, in this Appendix we consider a one-dimensional system but one would obtain the same results for higher dimension. Remarkably, for the broad family of systems described by Hamiltonian \eqref{hscale} the exact quantum dynamics of an arbitrary eigenstate under a modulation of the trapping frequency $\om(t)$ is described by a scaling law according to which the time-evolving state can be written in terms of the state at $t=0$ \cite{Gritsev10,delcampo11b}, 
\beqa
\label{psit}
\Psi_{\{m_i\}}\!\left(
z_1,\dots,z_\N,t\right)\!\!\!=\!
b^{-\frac{\N}{2}}\exp\left[i\frac{m\dot{b}}{2\hbar b}\sum_{i=1}^\N z_i^2-i\int_{0}^t\frac{E(\{m_i\})}{\hbar b(t')^2}dt'\right]\Psi_{\{m_i\}}\left(\frac{z_1}{b},\dots,\frac{z_\N}{b},t=0\!\right)\,,
\eeqa
where $E(\{m_i\})$ is the eigenvalue of \eqref{hscale} associated with the multi-index $\{m_i|i=1,\dots,\N\}$ defining a complete set of quantum numbers. 
Here,  $b=b(t)>0$ is the  scaling factor that can be obtained as a solution of the Ermakov differential equation
\beqa
\label{EPE}
\ddot{b}+\om(t)^2b=\om_0^2b^{-3}\ .
\eeqa
The  boundary conditions $b(0)=1$ and $\dot{b}(0)=0$ follow from the requirement for  the scaling law to reduce to the initial eigenstate at $t=0$ at the beginning of the process.
For the Calogero-Sutherland model, equation (\ref{psit}) generalizes the scaling law previously reported for the ground state ($m_i=0$ for all $i=1,\cdots,\N)$ in \cite{Sutherland98,delcampo15} 
but follows from the exact coherent states under scale-invariant driving found for a broad family of many-body systems in \cite{delcampo11b,delcampo13}.

In what follows, we shall give a general derivation of the relation between the nonadiabatic and the adiabatic mean energies given before the equation (\ref{qasteq}) in the main body of the article, and that holds for all quantum fluids described by Hamiltonian \eqref{hscale}. The scaling dynamics (\ref{psit}) is associated with the $SU(1,1)$ dynamical symmetry group \cite{Gambardella75}, shared by the familiar single-particle time-dependent harmonic oscillator \cite{Lohe09}.
We follow Lohe \cite{Lohe09} and describe the time-evolution in terms of the action of two spatial unitary transformations, elements of $SU(1,1)$. To this end, we introduce 
\beqa
T_{\rm dil}&=&\exp\left[-i\frac{\log b}{2\hbar}\sum_{i=1}^\N (z_ip_i+p_iz_i)\right],\\
T_z&=&\exp\left[i\frac{m\dot{b}}{2\hbar b}\sum_{i=1}^\N z_i^2\right],
\eeqa
to derive the invariant of motion 
\beqa
\mathit{I}=TH_0T^{\dag}=\sum_{i=1}^{\N}\left[\frac{b^2p_i^2}{2m}+\frac{1}{2}mz_{i}^{2}\left(\frac{\omega(0)^2}{b^2}+\dot{b}^2\right)
-\frac{b\dot{b}}{2}\ (z_ip_i+p_iz_i)\right]
+b^2\sum_{i<j}V(z_i-z_j)\ ,
\eeqa
where $T=T_zT_{\rm dil}$, and where $H_0$ denotes the Hamiltonian \eqref{hscale} at $t=0$.
This invariant is an instance of the many-body invariants known under scale-invariant driving, see Eq. (56) in \cite{DJD14}.
It has time-dependent eigenfunctions $\Phi_{\{m_i\}}\left(z_1,\dots,z_\N,t\right)=T\Psi_{\{m_i\}}\left(z_1,\dots,z_\N,0\right)$ and time-independent eigenvalues $E(\{m_k\})$. It also means that the time-evolving state \eqref{psit} can be written as 
\begin{equation}\label{psit2}
\Psi_{\{m_i\}}\!\left(z_1,\dots,z_\N,t\right)=e^{i\alpha_t}T\Psi_{\{m_i\}}\left(z_1,\dots,z_\N,0\right)\ ,
\end{equation}
where $\alpha_t=\int_{0}^t\frac{E(\{m_i\})}{\hbar b(t')^2}dt'$ is a time-dependent phase.    
As a function of it, Hamiltonian (\ref{hscale}) can be rewritten as
\beqa
H=\frac{1}{b^2}\mathit{I}+i\hbar\frac{\partial T}{\partial t}T^{\dag}\ ,
\eeqa 
where
\beqa\label{dTT}
i\hbar\frac{\partial T}{\partial t}T^{\dag}=\sum_{i=1}^{\N}\left[-\frac{1}{2}mz_{i}^{2}\left(\frac{\ddot{b}}{b}+\frac{\dot{b}^2}{b^2}\right)+\frac{\dot{b}}{2b}\ (z_ip_i+p_iz_i)\right]\ .
\eeqa
By \eqref{psit2} we find that
\beqa
\la\Psi_{\{m_i\}}(t)|H|\Psi_{\{m_i\}}(t)\ra&=&
\la\Psi_{\{m_i\}}(0)|T^\dagger H T|\Psi_{\{m_i\}}(0)\ra
=\la\Psi_{\{m_i\}}(0)|\left(\frac{H_0}{b^2}+i\hbar T^\dagger \frac{\partial T}{\partial t}\right)|\Psi_{\{m_i\}}(0)\ra\ ,
\eeqa
where we use
\begin{equation}\label{TdT}
i\hbar\ T^\dagger \frac{\partial T}{\partial t}=\sum_{i=1}^{\N}\left[\frac{1}{2}\left(\dot{b}^2-\ddot{b}b\right)z_i^2+\frac{\dot{b}}{2b}(z_ip_i+p_iz_i)\right]\ ,
\end{equation}
which is derived from equation (\ref{dTT}). 
As a result, the mean nonadiabatic energy is given by
\beqa
\la\Psi_{\{m_i\}}(t)|H|\Psi_{\{m_i\}}(t)\ra&=&
\frac{E(\{m_i\})}{b^2}-\frac{m}{2}\left(b\ddot{b}
-\dot{b}^2\right)\la\Psi_{\{m_i\}}(0)|\sum_{i=1}^\N z_i^2|\Psi_{\{m_i\}}(0)\ra+\frac{\dot{b}}{2b}\la\Psi_{\{m_i\}}(0)|\sum_{i=1}^\N (z_ip_i+p_iz_i)|\Psi_{\{m_i\}}(0)\ra\ .\nonumber\\
\eeqa
During time-evolution, the mean-energy of an initial thermal state
\beqa\label{rhoeq}
\rho=\sum_{\{m_i\}}p_{\{m_i\}}|\Psi_{\{m_i\}}(0)\ra\la\Psi_{\{m_i\}}(0)|\,,
\eeqa
with $p_{\{m_i\}}\geq0$ and $\sum_{\{m_i\}}p_{\{m_i\}}=1$ is simply given by
\beqa\label{MeanH}
\la H(t)\ra=\sum_{\{m_i\}}p_{\{m_i\}}\la\Psi_{\{m_i\}}(t)|H|\Psi_{\{m_i\}}(t)\ra \ .
\eeqa
Collecting all the previous equations holding for the general Hamiltonian \eqref{hscale}, the nonadiabatic mean-energy following a change of $\om(t)$ is found to be 
\beqa
\label{avenscaling}
 \la H(t)\ra&=&\frac{1}{b^2} \la H(0)\ra+\frac{\dot{b}}{2b}\sum_{i=1}^{\N}\la\{z_i,p_i\}(0)\ra+\frac{m}{2}(\dot{b}^2-b\ddot{b})\sum_{i=1}^{N}\la z_i^2(0)\ra\, ,
\eeqa
where $p_i$ is the momentum of the $i$-th particle and $\{z_i,p_i\}=z_ip_i+p_iz_i$.  
As a result of the underlying dynamical symmetry, 
it suffices to know the evolution of the scaling factor and the initial expectation values at $t=0$ of the mean energy, squeezing and position dispersion to characterize the nonadiabatic dynamics. Thanks to the thermal equilibrium property of the initial state, we can compute explicitly these quantities for the general Hamiltonian \eqref{hscale}. 

First, it is not difficult to show that at thermal equilibrium the mean squeezing vanishes 
\begin{equation}\label{meansqueezingvanishes}
\la\{z_i,p_i\}(0)\ra=0\, .
\end{equation}
Indeed, it suffices to rewrite the squeezing operator as 
$$2i\hbar\{z_i,p_i\}=[z_i^2,p_i^2]=[z_i^2,H]\,,$$ 
where we used the general form \eqref{hscale}. From \eqref{MeanH} it is clear that 
$$\la[z_i^2,H](0)\ra=0\ ,$$
which holds for any density matrix diagonal in the eigenbasis of $H$, and in particular,  for the thermal density matrix $\rho$ \eqref{rhoeq}.

Secondly, one can show that the initial mean position dispersion is proportional to the mean energy of the system,
\begin{equation}\label{meansquareformula}
\sum_i\la z_i^2(0)\ra = \frac{\la H(0)\ra}{m\omega_0^2}\,.
\end{equation}
Again, this formula is very general and holds for an initial thermal state. 
To derive \eqref{meansquareformula} we use the Hellmann-Feynman theorem
$\sum_i\la z_i^2(0)\ra= (m\omega_0)^{-1}\partial\la H(0)\ra/\partial\omega_0$.
Using the scaling invariance of the potential $V$, we have $V(\sqrt{\omega_0}z)=\omega_0^{-1}V(z)$. It follows that
$H(0)=\omega_0\tilde{H}(0)$ with $\tilde{H}(0)=\sum_{i=1}^N\left(-\frac{\hbar^2}{2m}\frac{\partial^2}{\partial q^2}+\frac{m}{2}q^2\right)+\sum_{i<j}V(q_i-q_j)$ where we rescaled the position as $q_i=\sqrt{\omega_0}z_i$.
Therefore, we find that $\partial\la H(0)\ra/\partial\omega_0=\omega_0^{-1}\la H(0)\ra$. 

Gathering equations \eqref{avenscaling}, \eqref{meansqueezingvanishes}, and \eqref{meansquareformula}, we obtain
\beqa
\la H(t)\ra=\frac{Q^{\ast}(t)}{b_{\rm ad}^2} \la H(0)\ra\,,\label{aven}
\eeqa
where $b_{\rm ad}=[\om_0/\om(t)]^{1/2}$ and where the nonadiabatic factor $Q^{\ast}(t)$ that accounts for the amount of energy excitations over the adiabatic dynamics is given by
\beqa\label{Qstar}
Q^{\ast}(t)=b_{\rm ad}^2\left[\frac{1}{2b(t)^{2}}+\frac{\om(t)^2 b(t)^2}{2\om_0^2}+\frac{\dot{b}(t)^2}{2\om_0^2}\right]\,.
\eeqa
Equations \eqref{aven} and \eqref{Qstar} give the general scaling law of the nonadiabatic mean energy of a class of systems governed by Hamiltonian \eqref{hscale}.  We further recognize
\beqa
\la H(t)\ra_{\rm ad }=\frac{1}{b_{\rm ad}^2} \la H(0)\ra\,
\eeqa
as the mean energy in the adiabatic limit. As a result, $Q^{\ast}(t)$ is simply the ratio between the nonadiabatic and adiabatic mean energies.


\subsection{Equivalence of the nonadiabatic factor under scale-invariant dynamics and the Husimi formula}\label{Husimi}

In this section we  prove that the nonadiabatic factor $Q^*(t)$ introduced in equation \eqref{Qstar} is equal to that introduced by Husimi for the single-particle time-dependent harmonic oscillator  $Q^*_{\text{H}}(t)$  \cite{Husimi53},
\begin{equation}\label{IdentityHCS}
Q^*_{\text{H}}(t)=Q^*(t)\ . 
\end{equation}
As discussed, the nonadiabatic factor $Q^*(t)$ provides the ratio between the nonadiabatic and adiabatic mean energies  for an infinite family of harmonically-confined quantum fluids following a variation of the trapping frequency \cite{delcampo11b}, whenever the dynamics is scale-invariant and the initial state at $t=0$ is a thermal state (with vanishing expectation value of the squeezing operator).
The nonadiabatic factor introduced by Husimi is defined as 
\begin{equation}\label{QH}
Q^*_{\text{H}}(t)=\frac{\left(\dot{G_1}^2+\omega(t)^2G_1(t)^2\right)+\omega_0^2\left(\dot{G_2}^2+\omega(t)^2G_2(t)^2\right)}{2\omega_0\omega(t)}\ , 
\end{equation}
where $G_1(t)$ and $G_2(t)$ are two fundamental solution of the classical harmonic oscillator equation $\ddot{x}(t)+\omega(t)^2x(t)=0$ that satisfy the following initial conditions:
\begin{subequations}\label{FundSol}
\begin{eqnarray}
 G_1(0)=1,\ \dot{G_1}(0)=0\ ,\label{FundSol1}\\
 G_2(0)=0,\ \dot{G_1}(0)=1\ \label{FundSol2}.
\end{eqnarray}
\end{subequations}

The Wronskian of these two functions $W_{G_1,G_2}(t)=\dot{G_1}(t)G_2(t)-G_1(t)\dot{G_2}(t)$ is shown to be constant
\begin{equation}\label{Wronskian}
W_{G_1,G_2}(t)=W_{G_1,G_2}(0)=-1\ . 
\end{equation}

As a preliminary result, we recall the derivation of a simple and useful formula previously derived in \cite{Pinney} for the scaling factor $b(t)$ solution of the Ermarkov equation \eqref{EPE},
\begin{equation}\label{FormulaSF}
b(t)=\sqrt{G_1(t)^2+\omega_0^2G_2(t)^2}\ . 
\end{equation}
To prove \eqref{FormulaSF}, we first compute the first time-derivative of $b(t)$
\begin{equation}
\label{FirstDer}
\dot{b}(t)=\frac{\dot{G_1}(t)G_1(t)+\omega_0^2 \dot{G_2}(t)G_2(t)}{\sqrt{G_1(t)^2+\omega_0^2G_2(t)^2}}= \frac{\dot{G_1}(t)G_1(t)+\omega_0^2 \dot{G_2}(t)G_2(t)}{b(t)}\ ,\\
\end{equation}
as well as  its second derivative 
\begin{align*}
\ddot{b}(t)&=\frac{\ddot{G_1}(t)G_1(t)+\omega_0^2 \ddot{G_2}(t)G_2(t)}{b(t)}+\frac{\dot{G_1}(t)^2+\omega_0^2 \dot{G_2}(t)^2}{b(t)}-\dot{b}(t)\frac{\dot{G_1}(t)G_1(t)+\omega_0^2 \dot{G_2}(t)G_2(t)}{b(t)^2}\ ,\\
&=-\omega(t)^2\frac{G_1(t)^2+\omega_0^2G_2^2}{b(t)}+\frac{(\dot{G_1}(t)^2+\omega_0 \dot{G_2}(t)^2)(G_1(t)^2+\omega_0^2G_2^2)}{b(t)^3}-\frac{(\dot{G_1}(t)G_1(t)+\omega_0^2 \dot{G_2}(t)G_2(t))^2}{b(t)^3}\ .
\end{align*}
The first term of the latter equation reduces to 
\begin{equation}\label{FirstTerm}
-\omega(t)^2 b(t)\ .
\end{equation} 
The second term is given by
\begin{align*}
\frac{\dot{G_1}(t)^2G_1(t)^2+\omega_0^2\dot{G_1}(t)^2G_2(t)^2+\omega_0^2\dot{G_2}(t)^2G_1(t)^2+\omega_0^4 \dot{G_2}(t)^2G_2(t)^2}{b(t)^3}\ ,
\end{align*}
while the third term gives
\begin{align*}
-\frac{\dot{G_1}(t)^2G_1(t)^2+2\omega_0^2\dot{G_1}(t)G_1(t)\dot{G_2}(t)G_2(t)
+\omega_0^4 \dot{G_2}(t)^2G_2(t)^2}{b(t)^3}\ .
\end{align*}
Thus, after adding these two terms, we find
\begin{align}\label{SecondThirdTerms}
\omega_0^2\frac{\left(\dot{G_1}(t)G_2(t)-\dot{G_2}(t)G_1(t)\right)^2}{b(t)^3}=\omega_0^2\frac{\left(W_{G_1,G_2}(t)\right)^2}{b(t)^3}=\frac{\omega_0^2}{b(t)^3}\ ,
\end{align}
where we used the property \eqref{Wronskian}. Gathering \eqref{FirstTerm} and \eqref{SecondThirdTerms} one gets the Ermarkov equation. One can easily check the consistency of the initial conditions using \eqref{FundSol} and \eqref{FormulaSF}-\eqref{FirstDer}. 

To prove the identity \eqref{IdentityHCS}, using  \eqref{FormulaSF} we rewrite \eqref{QH} and \eqref{Qstar} as
\begin{gather*}
Q^*_{\text{H}}(t)=\frac{\omega(t)^2G_1(t)^2+\omega_0^2\omega(t)^2G_2(t)^2}{2\omega_0\omega(t)}+\frac{\dot{G_1}^2+\omega_0^2\dot{G_2}^2}{2\omega_0\omega(t)}=\frac{\omega(t)b(t)}{2\omega_0} +\frac{\dot{G_1}^2+\omega_0^2\dot{G_2}^2}{2\omega_0\omega(t)}\ ,\\
Q^*(t)=\frac{\omega(t)b^2}{\omega_0}+\frac{1}{2\omega_0\omega(t)b^2}\left(\omega_0^2+\dot{b}^2b^2\right)\ , 
\end{gather*}
From \eqref{FirstDer} and  \eqref{Wronskian}, we have
\begin{gather*}
\dot{b}^2b^2=(\dot{G_1}G_1+\omega_0^2\dot{G_2}G_2)^2=\dot{G_1}^2G_1^2+2\omega_0^2\dot{G_1}G_1\dot{G_2}G_2+\omega_0^4\dot{G_2}^2G_2^2\ ,\\
\omega_0^2=\omega_0^2(W_{G_1,G_2}(t))^2
=\omega_0^2\left(\dot{G_1}^2G_2^2+\dot{G_2}^2G_1^2-2\dot{G_1}G_1\dot{G_2}G_2\right)\ .
\end{gather*}
Thus,
\begin{equation*}
\omega_0^2+\dot{b}^2b^2 = G_1^2\left(\dot{G_1}^2+\omega_0^2 G_2^2\right)+\omega_0^2\dot{G_2}^2\left(\dot{G_1}^2+\omega_0^2G_2^2\right) = b^2\left(\dot{G_1}^2+\omega_0^2\dot{G_2}^2\right)\ .
\end{equation*}
Gathering the last equalities we obtain the identity \eqref{IdentityHCS}.

\section{General results for the many-particle engine}\label{GeneralResults} 

\subsection{Preliminary}\label{GeneralResults1} 

Although the Calogero-Sutherland model described by the Hamiltonian in equation (\ref{eq:csm}) is a truly interacting many-body system, to account for its thermodynamics it is useful to exploit its mapping to effectively free harmonic oscillators \cite{Calogero71,Kawakami93,GP99,GP03}. 
Instead of using the set of quantum numbers $\{m_i|i=1,\dots,\N\}$, it is convenient to use the set of occupation numbers $\{n_j|j=1,\dots,\infty\}$ in each renormalized harmonic $j$-mode of energy $\epsilon_j=j\hbar\om$, with the identification $n_j=\sum_{i=1}^\N\delta_{j,m_i}$ where $\delta_{l,m}$ is the Kronecker delta. As shown in the main body of the article, the canonical partition function of the (bosonic) Calogero-Sutherland gas with interaction strength $\lambda$ is 
\begin{eqnarray}
Z_\N^{(\lambda)}=\!\!\!\!\!\!\!\sum_{\substack{\{n_j\}=1\\
                  \sum_{j=1}^{\infty}n_j=\N}}^\infty \!\!\! e^{-\beta E(\{n_j\})}=\prod_{k=1}^{\N}\frac{e^{-\beta\frac{\hbar\omega}{2}[1+ \lambda(\N-1)] }}{1-e^{-\beta k\hbar\omega}}\,,
\label{eq_zng}
\end{eqnarray}
where  $\{n_j\}$ is  the occupation number in the  $j$-th mode of energy $j\hbar\om$, satisfying the normalization $\sum_{j=1}^{\infty}n_k=\N$, and $E(\{n_j\})=\frac{\hbar\omega}{2}\N[1+ \lambda(\N-1)] +\sum_{j=1}^{\infty} j \hbar \omega n_{j}$.

It follows that the equilibrium mean energy, for a given frequency and temperature $(\omega,\beta)$, is given by
\begin{equation}\label{Energy}
\mathcal{E}_\N(\omega,\beta)=\frac{\hbar\omega \N}{2}\left[1+\lambda(\N-1)\right]+\sum_{k=1}^{\N}\frac{k\hbar\omega}{e^{k\hbar\beta\omega}-1}\ , 
\end{equation}
which can be read as
\beqa\label{EnergyDec} 
\mathcal{E}_\N(\omega,\beta)=\mathcal{E}^{(0)}(\omega)+\mathcal{E}^{(\text{th})}(\omega,\beta)\,,\eeqa
where the ground state energy $\mathcal{E}^{(0)}_\N(\omega)$ and the thermal energy $\mathcal{E}^{(\text{th})}_\N(\omega,\beta)$ correspond to the first and second term of \eqref{Energy}, respectively.
Depending on the value of $\sigma=\N\hbar\beta\omega$, three different regimes can be distinguished:
\begin{itemize}
 \item At very low temperature $\sigma\gg \N$, the mean energy is approximately given by the ground state energy
 \begin{equation}\label{VLT}
 \mathcal{E}_\N(\omega,\beta)\approx \mathcal{E}^{(0)}_\N(\omega) = \frac{\N\hbar\omega}{2}\left[1+\lambda(\N-1)\right]\ . 
 \end{equation}
 
 \item At high temperature $\sigma \ll 1$, we find corrections to the thermal energy due to quantum fluctuations
 \begin{equation}\label{LT}
 \mathcal{E}_\N(\omega,\beta)\approx \frac{\N}{\beta}\left[1+\frac{\sigma}{2}g_\N(\lambda-\frac{1}{2})+\sigma^2\frac{(\N+1)(2N+1)}{72N^2}\right]\ , 
 \end{equation}
 where $g_\N=(\N-1)\N^{-1}$ and we have disregarded  corrections of order $O(\sigma^3)$.   
 Formula \eqref{LT} is derived using  the fact that for $x<1$, $(e^x-1)^{-1}\approx\frac{1}{x}-\frac{1}{2}+\frac{x}{12}$  so that $$\mathcal{E}_\N(\omega,\beta) = \mathcal{E}^{(0)}_\N(\omega) + \frac{1}{\beta}\sum_{k=1}^{\N}\left(1-\frac{k\sigma}{2N}+\frac{k^2\sigma^2}{12N^2}\right)\ .$$ 
The sums can be carried out explicitly $\sum_{k=1}^{\N} k=\N(\N+1)/2$ and $\sum_{k=1}^{\N}k^2=\N(\N+1)(2N+1)/6$.

  \item For many-particles, a novel regime  absent in the single particle case is found for intermediate temperatures such that $1<\sigma\ll \N$ (which requires $\N$ to be large compared to $1$)
 \begin{equation}\label{Int}
 \mathcal{E}_\N(\omega,\beta)\approx \mathcal{E}_\N^{(0)}(\omega)+\frac{1}{\hbar\omega\beta^2}\int_{0}^{\sigma}ds\frac{s}{e^s-1}\approx\frac{\N}{\beta}\mu_\lambda(\sigma)\ ,
 \end{equation}
 where
 \begin{equation}\label{mu}
 \mu_\lambda(\sigma)=\frac{1}{\sigma}\int_{0}^{\sigma}ds\frac{s}{e^s-1}+\sigma\frac{\lambda}{2}\ .
 \end{equation}
To obtain the last equation, we estimated the Riemann sum in \eqref{Energy} by a Riemann integral since $\hbar\beta\omega\ll1$ (the corrections are bounded by $\N(\hbar\beta\omega)^2/2$ and so $\N$ can not be too large). 
The integral in \eqref{mu} can be found in closed form
$$\int_{0}^{\sigma}ds\frac{s}{e^s-1}=\frac{\pi^2}{6}+\sigma\log{(1-e^{-\sigma})}-\text{Li}_2(e^{-\sigma})\,,$$ 
in terms of the  standard polylogarithm function  $\text{Li}_n(z)=\sum_{j=1}^{\infty}j^{-n}z^j,\ n\geq 1$ \cite{Abramowitz}. The behavior of this function as a function of $\sigma$ for different values of $\lambda$ is shown in Fig. \ref{figSM:figure1}. We will see in the following sections  that this function plays an important role in the optimization of the total work.
\begin{figure}[t]
\begin{center}
\includegraphics[scale=0.8]{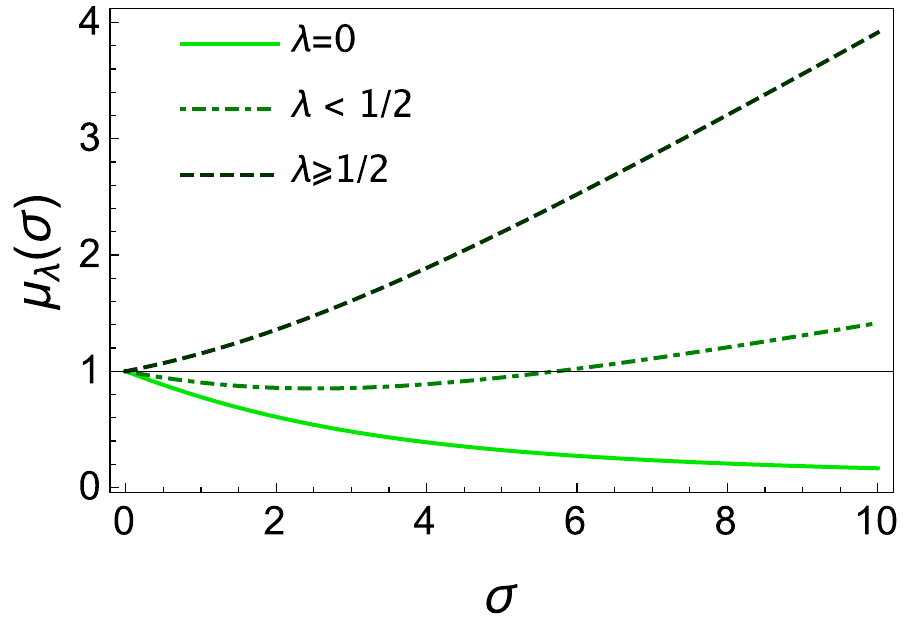}
\end{center}
\caption{\textbf{Variation of the function $\mu_{\lambda}(\sigma)$ defined by \eqref{mu} for different interaction strengths.} The straight continuous line gives the ordinate equal to $1$ which is crossed for a critical value of $\sigma$ for $0<\lambda<1/2$. For $\lambda=0$ the function is always less than $1$ while it is greater for $\lambda\geq 1/2$. }
\label{figSM:figure1}
\end{figure}

 \item The very high temperature limit, $\N\hbar\beta\omega\ll 1$, corresponds to the classical limit (quantum fluctuation are negligible in this regime) where the mean energy reads
 \begin{equation}\label{HT}
 \mathcal{E}_\N(\omega,\beta)\approx \frac{\N}{\beta}\ .
 \end{equation}
We note that to derive \eqref{HT} we use  formula \eqref{LT} where we neglect the quantum corrections $O(\hbar\beta\omega)$. 
\end{itemize}

Consider now a given protocol $\omega(t)$ for the compression stroke such that $\omega(0)=\omega_1,\ \omega(\tau)=\omega_2$ (similarly, for the compression stroke, let $\tilde{\omega}(t)$ satisfy $\tilde{\omega}(0)=\omega_2,\ \tilde{\omega}(\tau)=\omega_1$). We denote the nonadiabatic factor $Q_{AB}^*(\tau)$ (resp. $Q_{CD}^*(\tau)$) corresponding to the step $1$ of the Otto cycle $A\rightarrow B$ (resp. the step 3 $C\rightarrow D$), see formula \eqref{QH}. Let us assume that for any $\tau>0$ the nonadiabatic factor is continuously differentiable with respect to $\omega_1$ and $\omega_2$. We recall that for the adiabatic case, $Q_{AB}^*=Q_{CD}^*=1$ while in the nonadiabatic sudden-quench limit $Q_{AB}^*=Q_{CD}^*=\frac{\omega_1^2+\omega_2^2}{2\omega_1\omega_2}=\frac{x^2+1}{2x}$ with $x=\omega_1/\omega_2$. 

For the Otto cycle, the total work is usually defined as $W=\langle W_1\rangle+\langle W_3\rangle$, where $\langle W_1\rangle>0$ is the work corresponding to the compression ($\omega_1\rightarrow\omega_2>\omega_1$) and $\langle W_3\rangle<0$ is the work corresponding to the expansion ($\omega_2\rightarrow\omega_1<\omega_2$).
The engine produces work when $W$ is negative. Alternatively, we define $$\tilde{W}=-(\langle W_1\rangle+\langle W_3\rangle)\ .$$

We recall the equations given in the main body of the article: 
\begin{subequations}\label{WQWQ}
\begin{gather}
 \langle W_1\rangle = \langle H\rangle_B-\langle H\rangle_A = (Q_{AB}^*\frac{\omega_2}{\omega_1}-1)\langle H\rangle_A\ ,\label{WQ1} \\
 \langle W_3\rangle = \langle H\rangle_D-\langle H\rangle_C = (Q_{CD}^*\frac{\omega_1}{\omega_2}-1)\langle H\rangle_C\ , \label{WQ2}\\
 \langle \mathrm{Q}_2\rangle= \langle H\rangle_C-\langle H\rangle_B = \langle H\rangle_C - Q_{AB}^*\frac{\omega_2}{\omega_1}\langle H\rangle_A\ ,\label{WQ3}\\
 \langle \mathrm{Q}_4\rangle= \langle H\rangle_D-\langle H\rangle_A = Q_{CD}^*\frac{\omega_1}{\omega_2}\langle H\rangle_C - \langle H\rangle_A\ ,\label{WQ4}
\end{gather}
\end{subequations}
where $\langle H\rangle_A=\mathcal{E}_\N(\omega_1,\beta_{\rm c})$ and $\langle H\rangle_C=\mathcal{E}_\N(\omega_2,\beta_{\rm h})$ denote the energies at the equilibrium. 

As a consequence, the total work of the engine is given by
\begin{equation}\label{TotalWorkSM}
\tilde{W}=(1-Q_{AB}^*\frac{\omega_2}{\omega_1})\langle H\rangle_A+(1-Q_{CD}^*\frac{\omega_1}{\omega_2})\langle H\rangle_C\ ,
\end{equation}
which is also also equal to $\tilde{W}=\langle \mathrm{Q}_2\rangle+\left(\langle H\rangle_A-Q_{CD}^*\frac{\omega_1}{\omega_2}\langle H\rangle_C\right)$.
Then, the efficiency can be written in the following form
\begin{equation}\label{EfficiencySM}
\eta=\frac{\tilde{W}}{\langle \mathrm{Q}_2\rangle}=1-\frac{\omega_1}{\omega_2}\frac{Q_{CD}^* \langle H\rangle_C-\frac{\omega_2}{\omega_1}\langle H\rangle_A}{\langle H\rangle_C-Q_{AB}^*\frac{\omega_2}{\omega_1}\langle H\rangle_A}\ .
\end{equation}
In the adiabatic case ($Q^*=1$) the efficiency reduces to the Otto efficiency
\begin{equation}\label{EfficiencyAd}
\eta_{\text{ad}}=1-\frac{\omega_1}{\omega_2}\ ,
\end{equation}
which is independent of $\N$ and $\lambda$. However, for an adiabatic driving the total work \eqref{TotalWorkSM} depends on $\N$ (and not on $\lambda$ as we shall see in the section \ref{GeneralResults2}). Thus, one can expect that the frequency ratio $\omega_1/\omega_2$ at which the work output is maximum depends on $\N$. The efficiency at optimal work inherits a nontrivial dependence on $\N$, as discussed in  the section \ref{Adiabatic}.

For a finite-time protocol, the nonadiabatic factors $Q^*_{AB(CD)}\geq 1$ and  the efficiency is bounded from above by the adiabatic Otto efficiency, which constitutes a universal limit as it does not depend on $N,\lambda$  or the specific driving protocol. This bound can be too conservative for finite-time thermodynamics. For example, for a sudden quench driving the efficiency can actually be upper bounded by {\it half} the adiabatic Otto efficiency. Hence, it is convenient to derive a tighter upper bound for the efficiency. To this end, let us rewrite the efficiency as follows
\begin{equation*}
\eta=1-Q_{CD}^*\frac{\omega_1}{\omega_2}\frac{ \langle H\rangle_C-(Q_{CD}^*)^{-1}\frac{\omega_2}{\omega_1}\langle H\rangle_A}{\langle H\rangle_C-Q_{AB}^*\frac{\omega_2}{\omega_1}\langle H\rangle_A}\ .
\end{equation*}
Since $(Q_{CD}^*)^{-1}\leq 1$ and $Q_{AB}^*\geq 1$, we find the following upper bound
\begin{equation}\label{UB}
\eta\leq 1-Q_{CD}^*\frac{\omega_1}{\omega_2}\ ,
\end{equation}
where $Q_{CD}^*\frac{\omega_1}{\omega_2}$ is nothing but $\frac{\langle H\rangle_D}{\langle H\rangle_C}$, see \eqref{WQ4}.
Notice that this limit depends only on the factor $Q_{CD}^*$ and not on $Q_{AB}^*$. An upper bound depending on $Q_{AB}^*$ can as well be obtained,   $\eta\leq 1-(Q_{AB}^*)^{-1}\frac{\omega_1}{\omega_2}=1-\frac{\langle H\rangle_A}{\langle H\rangle_B}$, but this is not interesting as it is weaker than the (adiabatic) Otto efficiency. 

In what follows, we shall provide a  derivation of the efficiency at optimal output power $P=\tilde{W}/\tau$, where $\tau$ is the running time of a cycle and $\tilde{W}$ is the absolute value of the output work per cycle. To this end, we first find the optimal frequency ratio corresponding to the maximal value of the total work per cycle using the exact expression given by equation \eqref{TotalWorkSM}, and then we insert this optimal value in equation \eqref{EfficiencySM} to obtain the efficiency at optimal output power. 

\subsection{Effect of $\lambda$ on the total work and on the efficiency}\label{GeneralResults2} 

In this section we discuss  the effect of the interaction strength $\lambda$ on the total work and the efficiency, for arbitrary dynamics. 
We consider a finite-time protocol for the Otto cycle as described in the previous section. \\

\textbf{General statement:}
\begin{enumerate}[label=(\roman*)]
\item The total work (as well as the optimal work) and the efficiency are two monotonically decreasing functions of $\lambda$.
\item In the adiabatic limit the total work and the efficiency do not depend on $\lambda$.   
\end{enumerate}

We prove this important statement. We introduced $\lambda$ as a superscript, explicitly.
From the total work \eqref{WQ1}-\eqref{WQ4}, one can easily show that
\begin{align*}
\tilde{W}^\lambda&=\tilde{W}^0+\frac{\hbar}{2}\left(\omega_1-Q_{AB}^*\omega_2\right)C_{\N,\lambda}+\frac{\hbar}{2}\left(\omega_2-Q_{CD}^*\omega_1\right)C_{\N,\lambda}\ ,\\
&=\tilde{W}^0+\frac{\hbar}{2}\left[\omega_1-\omega_2-(Q_{AB}^*-1)\omega_2\right)C_{\N,\lambda}+\frac{\hbar}{2}\left(\omega_2-\omega_1-(Q_{CD}^*-1)\omega_1\right]C_{\N,\lambda}\ ,\\
&=\tilde{W}^0+\frac{\hbar}{2}\left[(1-Q_{AB}^*)\omega_2+(1-Q_{CD}^*)\omega_1\right]C_{\N,\lambda}\ ,
\end{align*} 
where $C_{\N,\lambda}=\N(\N+1)\lambda$ and $\tilde{W}^0$ is the total work for $\lambda=0$.
Knowing that $Q_{AB}^*\geq 1$ and $Q_{CD}^*\geq 1$, it follows that $\partial_\lambda \tilde{W}^\lambda\leq 0$. In addition, for the adiabatic case one finds $\tilde{W}^\lambda=\tilde{W}^0$ (since $Q^*=1$) and the total work does not depends on $\lambda$.  
Similarly, one can show that $\langle \mathrm{Q}_2\rangle^\lambda=\langle \mathrm{Q}_2\rangle^0+\frac{\hbar\omega_2}{2}(1-Q_{AB}^*)C_{\N,\lambda}$  so  $\partial_\lambda \langle \mathrm{Q}_2\rangle^\lambda\leq 0$. 
In regard to the efficiency:
\begin{equation*}
\eta^\lambda=\frac{\tilde{W}^\lambda}{\langle \mathrm{Q}_2\rangle^\lambda}=\frac{\tilde{W}^0+\frac{\hbar}{2}\left[(1-Q_{AB}^*)\omega_2+(1-Q_{CD}^*)\omega_1\right]C_{\N,\lambda}}{\langle \mathrm{Q}_2\rangle^0+\frac{\hbar\omega_2}{2}(1-Q_{AB}^*)C_{\N,\Lambda}}\ .
\end{equation*}
The first derivative of the above quantity with respect to $\lambda$ is given by
\begin{align*}
\partial_\lambda\eta^\lambda&=\frac{\frac{\hbar}{2}\left[(1-Q_{AB}^*)\omega_2+(1-Q_{CD}^*)\omega_1\right]\N(\N+1)}{\langle \mathrm{Q}_2\rangle^\lambda}-\frac{\hbar\omega_2}{2}(1-Q_{AB}^*)\N(\N+1)\frac{\tilde{W}^\lambda}{(\langle \mathrm{Q}_2\rangle^\lambda)^2}\ ,\\
&=\frac{\hbar \omega_1 (1-Q_{CD}^*) \N(\N+1)}{2 \langle \mathrm{Q}_2\rangle^\lambda}+\frac{\hbar\omega_2(1-Q_{CD}^*)\N(\N+1)}{2\langle \mathrm{Q}_2\rangle^\lambda}\left(1-\eta^\lambda\right)\leq 0\ .
\end{align*}
It follows that for the adiabatic case ($Q_{AB}^*=Q_{CD}^*=1$) the efficiency does not depends on $\lambda$, i.e., $\eta^\lambda=\eta^0$. 

Next we  show that the optimal work is a monotonically increasing function of $\lambda$. For the sake of simplicity, we introduce the parameter $x=\omega_1/\omega_2$. We have $\tilde{W}^\lambda(x)=\tilde{W}^0(x)+\frac{\hbar\omega_1}{2}\left[(1-Q_{AB}^*)\frac{1}{x}+(1-Q_{CD}^*)\right]C_{\N,\lambda}$. We proved that $\forall x\in[0,1]$, $\partial_\lambda \tilde{W}^\lambda\leq 0$ which means that $\forall x\in[0,1]$ and $\forall \lambda,\lambda'$ such that $\lambda\leq\lambda'$, $\tilde{W}^\lambda(x)\geq \tilde{W}^{\lambda'}(x)$, and so $\forall x\in[0,1],\ \mathrm{max}_{y\in[0,1]}\tilde{W}^\lambda(y)\geq \tilde{W}^\lambda(x)$, thus if there exists $x^{\lambda}_{\rm opt}$ and $x^{\lambda'}_{\rm opt}$ such that $\tilde{W}^\lambda(x^\lambda_{\rm opt})=\mathrm{max}_{y\in[0,1]}\tilde{W}^\lambda(y)<\infty$ and $\tilde{W}^{\lambda'}(x^{\lambda'}_{\rm opt})=\mathrm{max}_{x\in[0,1]}\tilde{W}^{\lambda'}(x)<\infty$, then $\tilde{W}^{\lambda}_{\rm opt}\geq \tilde{W}^{\lambda'}_{\rm opt}$.

\subsection{Quantum Supremacy under identical resources}\label{AppendixQS}

We introduce two parameters 
\begin{gather}\label{varxa}
x=\frac{\omega_1}{\omega_2},\ a=\frac{\beta_{\rm h}}{\beta_{\rm c}}\ ,
\end{gather}
with $a<1,\ x<1$. 

To quantify the boost in the performance of the QHE resulting from  many-particle effects, we introduce the following quantities
\begin{subequations}\label{ratioSM}
\begin{equation}
r_{\N,\lambda}(x,a)\equiv \frac{W_{\N,\lambda}(x,a)}{\N\times W_{1}(x,a)}\ ,\label{ratioSM1} 
\end{equation}
\begin{equation}
\rho_{\N,\lambda}(x,a)\equiv \frac{\eta_{\N,\lambda}(x,a)}{\eta_{1}(x,a)}\ .\label{ratioSM2}
\end{equation}
\end{subequations}
The first one is the  the ratio between the total output work $W_{\N,\lambda}(x,a)$  of a many-particle QHE and the corresponding value of $\N$ independent  single-particle QHEs given by $\N\times W_{1}(x,a)$
under the same resources, that is,  for the same values of $\N,\lambda,\beta_{\rm c,h},\omega_{1,2}$. 
The ratio $\rho_{\N,\lambda}(x,a)$ between the corresponding efficiencies is defined analogously. 
We note that the ratio between output power is equal to $\rho_{\N,\lambda}(x,a)$ as the cycle time is the same for both systems.  

Assuming that the temperature of the cold reservoir is high $\hbar\omega_1\beta_{\rm c}\ll 1$ the total output work as well as the efficiency of a single-particle QHE  are dominated by the classical values, see equation \eqref{LT} (taking $\N=1$) and \eqref{etaSQ:sigmasmall}, as well as the Appendix \ref{SuddenQuench} for a sudden quench protocol.  
This is no longer the case for the many-particle QHE as the Planck constant is scaled up by the number of particles  and quantum effects become comparable with the classical ones, see equation \eqref{Int}. In this large-$\N$ regime, we  observe numerically that both ratios in equation \eqref{ratioSM} are greater than one. In other words, the performance of the many-particle QHE can surpass that of an ensemble of single-particle QHEs with the same resources. We call this phenomenon quantum supremacy. In the main body of this article we look at an alternative definition of quantum supremacy considering the many-particle QHE and the series of single-particle QHEs with the same number of particles $\N$, temperatures $\beta_{\rm c,h}$, trap frequency $\omega_1$, and interaction strength $\lambda$ but with a frequency  ratio $x=\omega_1/\omega_2$ that corresponds to the value optimizing the output power of the respective engines. 

Here we show that for the same resources (i.e. when the value of $\omega_2$ is also fixed) the many-particle efficiency is always greater than the classical single-particle efficiency for a given range of the interaction strength $\lambda$. While no such general statement  holds for the total output work, regimes exhibiting quantum supremacy can be found, showing the robustness of this phenomenon.
We consider $\hbar\omega_1\beta_{\rm c}\ll 1$ and $\sigma_{\rm c},\sigma_{\rm h}>1$, meaning that $\N$ is large compared to $1$. Then, from equations \eqref{Int}, \eqref{WQWQ}, \eqref{TotalWorkSM}, and \eqref{EfficiencySM} we find 
\begin{subequations}\label{WQS}
\begin{equation}
W_{\N,\lambda}=\frac{\N}{\beta_{\rm c}}\left[\left(1-\frac{Q_{AB}^\ast}{x}\right)\mu_{\lambda}(\sigma_{\rm c})+\left(1-xQ_{CD}^\ast\right)\frac{1}{a}\mu_{\lambda}(\sigma_{\rm h})\right]\ ,
\end{equation}
\begin{equation}
\N \times W_1 = \frac{\N}{\beta_{\rm c}}\left[\left(1-\frac{Q_{AB}^\ast}{x}\right)+\left(1-xQ_{CD}^\ast\right)\frac{1}{a}\right]\ ,
\end{equation}
\end{subequations}
and
\begin{subequations}\label{EffQS}
\begin{equation}
\eta_{\N,\lambda}=\frac{\left(1-\frac{Q_{AB}^\ast}{x}\right)\mu_{\lambda}(\sigma_{\rm c})+\left(1-xQ_{CD}^\ast\right)\frac{1}{a}\mu_{\lambda}(\sigma_{\rm h})}{\frac{1}{a}\mu_{\lambda}(\sigma_{\rm h})-\frac{Q_{AB}^\ast}{x}\mu_{\lambda}(\sigma_{\rm c})}\ ,
\end{equation}
\begin{equation}
\eta_{1}=\frac{\left(1-\frac{Q_{AB}^\ast}{x}\right)+\left(1-xQ_{CD}^\ast\right)\frac{1}{a}}{\frac{1}{a}-\frac{Q_{AB}^\ast}{x}}\ .
\end{equation}
\end{subequations}
As a result,
\begin{subequations}\label{RatioQS}
\begin{equation}\label{RatioQSr}
r_{\N,\lambda}(x,a)=\mu_{\lambda}(\sigma_{\rm c})\frac{W_1(x,a f_{\lambda})}{W_1(x,a)}\ ,
\end{equation}
\begin{equation}\label{RatioQSrho}
\rho_{\N,\lambda}(x,a)=\frac{\eta_1(x,a f_{\lambda})}{\eta_1(x,a)}\ ,
\end{equation}
\end{subequations}
where 
\begin{equation}\label{flambda}
f_{\lambda}=\frac{\mu_{\lambda}(\sigma_{\rm c})}{\mu_{\lambda}(\sigma_{\rm h})}\ ,
\end{equation}
with $\sigma_{\rm h}=\frac{a}{x}\sigma_{\rm c}<\sigma_{\rm c}$.  
In equations \eqref{RatioQSr} and \eqref{RatioQSrho} we see that the total work and the efficiency of a many-particle QHE can be written by rescaling the ratio of temperatures $a=\beta_{\rm h}/\beta_{\rm c}$ in the single-particle  quantities $W_1$ and $\eta_1$.  One finds that
\begin{subequations}\label{DerivativeWeta}
\begin{equation}\label{DerivativeW}
\frac{\partial}{\partial a}W_1(x,a)=-\frac{(1-x Q_{CD}^\ast)}{a^2} < 0,\ \mathrm{for\ all}\ a,x\ ,
\end{equation}
\begin{equation}\label{Derivativeeta}
\frac{\partial}{\partial a}\eta_1(x,a)=-\frac{(Q_{AB}^\ast Q_{CD}^\ast-1)x^2}{(x-a Q_{AB}^\ast)^2} < 0,\ \mathrm{for\ all}\ a,x\ .
\end{equation}
\end{subequations} 
From  equations \eqref{Derivativeeta} and \eqref{flambda} we find that the  efficiency ratio \eqref{RatioQSrho} exceeds unity if the function $\mu_{\lambda}(\sigma)$ defined by \eqref{mu} is decreasing. We know that it is monotonically decreasing for $\lambda=0$ and monotonically increasing for $\lambda\geq 1/2$. This means that for the ideal Bose gas the many-particle efficiency is always greater than the corresponding value for an ensemble of single-particle QHEs in the classical regime. For $0<\lambda<1/2$, the function $\mu_{\lambda}(\sigma)$ decreases if $\sigma$ is less than a critical value that increases with $\lambda$, see Fig. \ref{figSM:figure1}. Consequently for $\sigma\leq \sigma_{\rm c}$, $\mu_\lambda(\sigma)$ is a decreasing function of $\sigma$ if the interaction strength is below a critical value $\lambda_c$ which is typically less than $0.2$ (we obtained this estimate after numerical computations). Thus the ratio $\rho_{\N,\lambda}(x,a)$ is greater than $1$ for $\lambda\leq\lambda_c$ and less than $1$ otherwise. In other words, many-particle quantum effects enhance the efficiency of a QHE for weak interactions. 

Given our focus on quantum supremacy, we would also like to show the condition to improve the output power of the engine, i.e., for a ratio \eqref{RatioQSr} greater than unity. From equation \eqref{DerivativeW} and a similar argument to that used above, we know that the ratio between $W_1(x,a f_\lambda)$ and $W_1(x,a)$ is greater than $1$ for $\lambda=0$ and for $0<\lambda\leq \lambda_c$, where $\lambda_c$ is the same critical value discussed in the previous paragraph. However, in equation \eqref{RatioQSr} we find that this ratio is multiplied by a factor $\sigma_\lambda(\sigma_{\rm c})$ which is less than $1$ for $\lambda\leq \lambda_c$, making the analysis more complex. Yet, we find that for $\sigma_{\rm c}$ not too large  (typically $\sigma_{\rm c}\leq 5-10$, depending on the values of $a$ and $x$) the output power of a many-particle QHE surpasses the corresponding value of an ensemble of single-particle QHEs, exhibiting quantum supremacy.

\section{Total work and efficiency in the adiabatic limit}\label{Adiabatic} 

Given that the nonadiabatic factor reduces to unity $Q^*=1$  under slow driving, using  \eqref{Energy} the adiabatic heat can explicitly written as
\begin{equation}
 \langle\mathrm{Q}_2\rangle = \mathcal{E}_\N(\omega_2,\beta_{\rm h})- \frac{\omega_2}{\omega_1} \mathcal{E}_\N(\omega_1,\beta_{\rm c})= 
 \sum_{k=1}^{\N}k\hbar\omega_2\left(\frac{1}{e^{k\hbar\beta_{\rm h}\omega_2}-1}-\frac{1}{e^{k\hbar\beta_{\rm c}\omega_1}-1}\right)\ .
\end{equation}
It is clear that during the  hot isochore the working medium absorbs heat $\langle \mathrm{Q}_2\rangle\geq0$, this leads to the following condition: 
\begin{equation}\label{ConditionQ2}
\beta_{\rm h}\omega_2\leq\beta_{\rm c}\omega_1\ \mathrm{or}\ a\leq x\ .
\end{equation}
Under nonadiabatic driving, $Q^*$ is generally greater than $1$ so $\langle \mathrm{Q}_2\rangle>0$ obeys a stronger condition. We will discuss this condition for the sudden quench in the next section. \\

In this section, we derive the formulas for the optimal work and efficiency at optimal work for the adiabatic case. First, we set $Q^*=1$ in Eqs. \eqref{WQ1}-\eqref{WQ4} and \eqref{EnergyDec} to find
\begin{equation}\label{Wad11}
\tilde{W}_{\text{ad}}(x)=\left(1-\frac{1}{x}\right)\mathcal{E}^{(\text{th})}(\omega_1,\beta_{\rm c})+(1-x)\mathcal{E}^{(\text{th})}(\frac{\omega_1}{x},a\beta_{\rm c}) \ ,
\end{equation}
that does not depends on $\lambda$ as we proved in the previous section. Also, we note that $\tilde{W}_{\text{ad}}(x)$ only depends on the thermal contribution because $\langle H\rangle_A=\frac{\N\hbar\omega_1}{2}+\mathcal{E}^{(\text{th})}(\omega_1,\beta_{\rm c})$ and $\langle H\rangle_C=\frac{\N\hbar\omega_2}{2}+\mathcal{E}^{(\text{th})}(\omega_1/x,a\beta_{\rm c})$.
 
Assuming $\sigma_h=\N\beta_{\rm h}\hbar\omega_2\ll 1$ we obtain
$$\tilde{W}_{\text{ad}}=\left(1-\frac{1}{x}\right)E_{\N,1/2}(\omega_1,\beta_{\rm c})+(1-x)\frac{\N}{a\beta_{\rm c}}\ ,$$
where $E_{\N,\lambda}(\omega_1,\beta_{\rm c})=\langle H\rangle_A$ and so $E_{\N,1/2}(\omega_1,\beta_{\rm c})=\mathcal{E}^{(\text{th})}(\omega_1,\beta_{\rm c})+\frac{\hbar\omega_1}{4}\N(\N+1)$ consistently with \eqref{Wad11} and \eqref{LT}. After differentiating the total work with respect to the frequency ratio $x=\omega_{1}/\omega_{2}$, we find the following solution of the equation $\partial_x \tilde{W}=0$:
\begin{equation}\label{XoptAd}
\overline{x}_{\text{ad}}=\sqrt{\frac{a\beta_{\rm c} E_{\N,1/2}(\omega_1,\beta_{\rm c})}{\N}}\ .
\end{equation}
The optimal work and the efficiency at the optimal work are then given by
\begin{subequations}\label{WoptAd}
\begin{eqnarray}
\tilde{W}_{\text{ad}}^{(\N)}&=& \frac{\N}{a\beta_{\rm c}}\left(1-\sqrt{\alpha_{\text{ad}}}\right)^2+O(\sigma_{\rm c}^2)\ ,\label{WoptAd1}\\
\eta_{\text{ad}}^{(\N)}&=& 1-\sqrt{\alpha_{\text{ad}}}+O(\sigma_{\rm c}^2)\ ,
\label{WoptAd2}
\end{eqnarray}
\end{subequations}

where $\alpha_{\text{ad}}=a\beta_{\rm c} E_{\N,1/2}(\omega_1,\beta_{\rm c})/\N$ and $\sigma_{\rm c}=\N\beta_{\rm c}\hbar\omega_1$.

\begin{figure}[t]
\begin{center}
\includegraphics[scale=0.8]{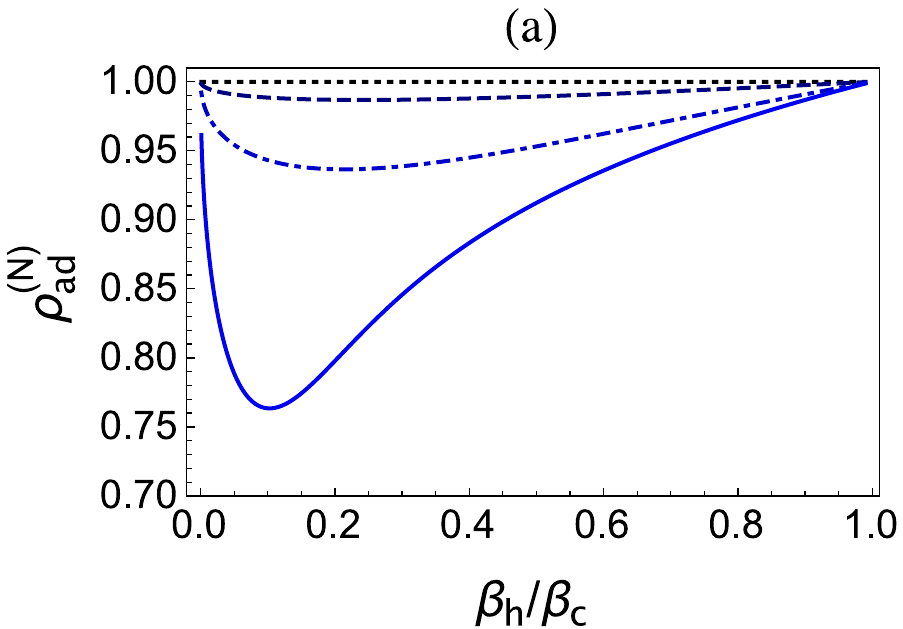}\includegraphics[scale=0.8]{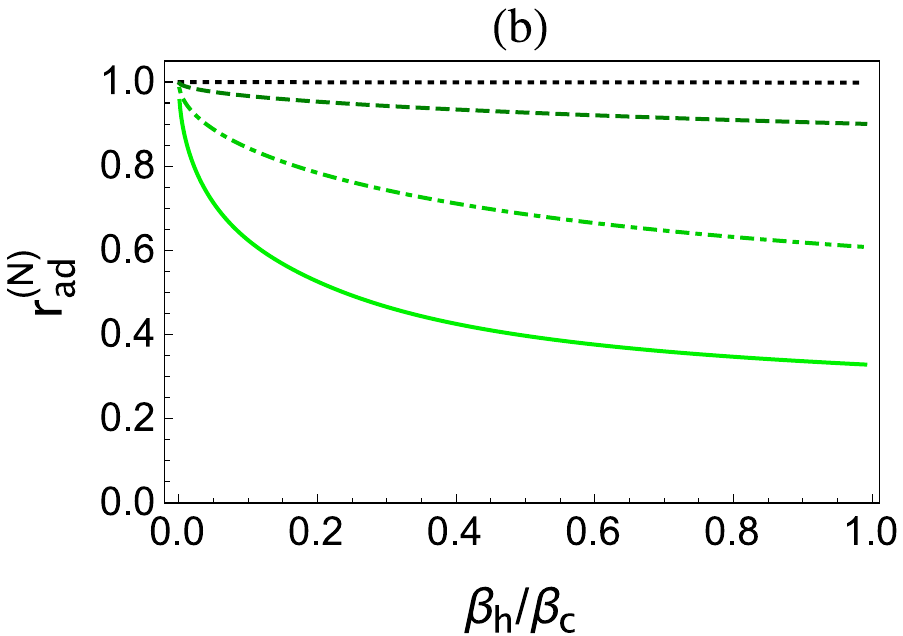}
\end{center}
\caption{\textbf{Performance of an adiabatic many-particle QHE compared to $\N$ single-particle QHE in series.} 
The adiabatic efficiency at optimal output power of a $\N$-particle QHE normalized by the single-particle value (a), and the adiabatic optimal output power divided by $\N$ times the single-particle value (b) (see equation \eqref{ratioSM}) are displayed for $\N=20,200,500$ and $2,000$ from top to bottom,
under an adiabatic driving with $\beta_{\rm c}=0.01/(\hbar\om_1)$. 
In the adiabatic limit, the efficiency becomes independent of the interaction strength $\lambda$ and many-particle quantum effects are detrimental.}
\label{figSM:figure2}
\end{figure}

Using the asymptotic expansion of the thermal energy, one can easily find the asymptotic expressions for the optimal work and the corresponding efficiency in different regimes distinguished by the value of $\sigma_{\rm c}$:
\begin{itemize}

\item For $\sigma_{\rm c}\gg N$, $\mathcal{E}^{(\text{th})}\approx 0$ so $E_{\N,1/2}(\omega_1,\beta_{\rm c})\approx \frac{\hbar\omega_1}{4}\N(\N+1)$ and then $\alpha_{\text{ad}}\approx \frac{\hbar\omega_1\beta_{\rm h} (\N+1))}{4}$ which gives
\begin{subequations}\label{OptWorkEffAdVLT}
\begin{eqnarray}
\tilde{W}^{(\N)}_{\text{ad}}&\approx & \frac{\N}{a\beta_{\rm c}}\left(1-\sqrt{\frac{\hbar\omega_1\beta_{\rm h} (\N+1)}{4}}\right)^2\ ,\label{OptWorkEffAdVLT1}\\
\eta_{\text{ad}}^{(\N)}&\approx & 1-\sqrt{\frac{\hbar\omega_1\beta_{\rm h} (\N+1)}{4}}\ .\label{OptWorkEffAdVLT2}
\end{eqnarray}
\end{subequations} 

\item For $\sigma_{\rm c}\ll 1$, $E_{\N,1/2}(\omega_1,\beta_{\rm c})\approx \frac{\N}{\beta_{\rm c}}\left[1+O(\sigma_{\rm c}^2)\right]$ and then $\alpha_{\text{ad}}\approx a\left[1+O(\sigma_{\rm c}^2)\right]$ leading to

\begin{subequations}\label{OptWorkEffAdHT}
\begin{eqnarray}
\tilde{W}_{\text{ad}}^{(\N)}&\approx &\frac{\N}{a\beta_{\rm c}}\left[\left(1-\sqrt{a}\right)^2+O(\sigma_{\rm c}^2)\right]\ ,\label{OptWorkEffAdHT1}\\
\eta_{\text{ad}}^{(\N)}&\approx & 1-\sqrt{a}+O(\sigma_{\rm c}^2)\ .\label{OptWorkEffAdHT2}
\end{eqnarray}
\end{subequations}

\item For $\N>1$, we consider an intermediate regime corresponding to a large temperature $\sigma_{\rm c}\ll N$ but a relatively small temperature per particle $\sigma_{\rm c} \geq 1$, which means that the particle number is large, $\N\gg1$ but keeping $\sigma_{\rm h}\ll 1$ (i.e., $\beta_2\ll\beta_1$). Using \eqref{Int} we find that
$E_{\N,1/2}(\omega_1,\beta_{\rm c})\approx \frac{\N}{\beta_{\rm c}}\mu_{1/2}(\sigma_{\rm c})$
where $\sigma_{\rm c}=\N\hbar\beta_{\rm c}\omega_1$, and $\mu_{1/2}(\sigma_{\rm c})$ is defined by \eqref{mu}. As a result, by \eqref{XoptAd} and \eqref{WoptAd},
\begin{subequations}
\begin{eqnarray}
\tilde{W}_{\text{ad}}^{(\N)} &\approx & \frac{\N}{2a\beta_{\rm c}}\left(1-\sqrt{\mu_{1/2}(\sigma_{\rm c}) a}\right)^2\label{OptWorkEffAdInt1},\\ 
\eta_{\text{ad}}^{(\N)} &\approx & 1-\sqrt{\mu_{1/2}(\sigma_{\rm c}) a} \ ,\label{OptWorkEffAdInt2}
\end{eqnarray}
\end{subequations}
where $\mu_{1/2}(\sigma_{\rm c})=\mu(\sigma_{\rm c})$ is greater than unity and increases monotonically with $\sigma_{\rm c}$, as shown in Fig. \ref{figSM:figure1}. Consequently, in this regime the efficiency is lesser than the Curzon-Ahlborn efficiency $1-\sqrt{a}$ associated with a classical heat engine operated in the adiabatic limit \cite{CA75}, see Fig. \ref{figSM:figure2} as well as Fig. \ref{figure2}. We emphasize that equations \eqref{OptWorkEffAdInt1} and \eqref{OptWorkEffAdInt2} are valid only for $a\ll1$. 
\end{itemize}

\section{Total work and efficiency for the sudden quench}\label{SuddenQuench} 

\subsection{Preliminary}

For the sudden-quench driving between $\om_1$ and $\om_2$ the nonadiabatic factor $Q^*$ takes the same value in both the compression and expansion strokes and is given by
\begin{equation}\label{NonAdFactorSQ}
Q^*=\frac{\omega_1^2+\omega_2^2}{2\omega_1\omega_2}\ . 
\end{equation}

We recall the condition over $a=\beta_{\rm h}/\beta_{\rm c}$ and $x=\omega_1/\omega_2$ \eqref{ConditionQ2} which can be read as $a\leq x$. One can derive another condition which is more useful for small $x$:
\begin{equation}\label{ConditionQ2nonad}
\frac{a}{x}\leq 2\,x,\ \mathrm{for}\ x\ll1\ . 
\end{equation}
To derive this condition we use \eqref{WQ2} to show that $\langle \mathrm{Q}_2\rangle=\frac{\omega_2}{2}\N(1+\lambda(\N-1))(1-Q^*)+\sum_{k=1}^{\N}k\hbar\omega_2\left(\frac{1}{e^{k\hbar\beta_{\rm h}\omega_2}-1}-\frac{Q^*}{e^{k\hbar\beta_{\rm c}\omega_1}-1}\right)$. We must have $\langle \mathrm{Q}_2\rangle\geq0$, and the sum in the previous equation must be positive since the first term is negative in the adiabatic case ($Q^*\geq1$). Using $x\leq e^x-1\leq 2x$ for $x\leq1$, one can easily show that an upper bound to the sum is positive if $\frac{\beta_{\rm c}\omega_1}{\beta_{\rm h}\omega_2}\geq2Q^*$ which gives \eqref{ConditionQ2nonad} for small $x$. This condition is weak in the sense that it does not ensure that $\langle \mathrm{Q}_2\rangle$ is positive. However, we know that if this condition is not satisfied then $\langle \mathrm{Q}_2\rangle$ is negative.\\

Using \eqref{TotalWorkSM} and \eqref{NonAdFactorSQ}, we find
\begin{equation}\label{TotalWorkxa}
\tilde{W}_{\text{sq}} = 
\left(\frac{x^2-1}{2x^2}\right)\langle H\rangle_A + \left(\frac{1-x^2}{2}\right)\langle H\rangle_C\ ,
\end{equation}
where $\langle H\rangle_A=\mathcal{E}_\N(\omega_1,\beta_{\rm c})$ and $\langle H\rangle_C=\mathcal{E}_\N(\omega_1/x,a\beta_{\rm c})$.

\subsection{Optimal output work and efficiency}

We want to find a condition on $x$ so that the work is maximal. To this end, we solve this equation
\begin{equation}\label{OptWorkeq}
\partial_x \tilde{W}_{\text{sq}}=0\ ,
\end{equation}
which gives the maximal work since the total work is a concave function of $x$, see \eqref{TotalWorkxa}.  

By \eqref{LT}, in the  high-temperature limit of the hot reservoir $\sigma_{\rm h}=N\hbar\beta_{\rm h}\omega_2\ll 1$, the total work is given by
\begin{equation}\label{TotalWorkHT}
\tilde{W}_{\text{sq}} = \left(\frac{x^2-1}{2x^2}\right)\langle H\rangle_A + \left(\frac{1-x^2}{2}\right)\frac{\N}{a\beta_{\rm c}}\left(1+\frac{a}{2x} g_\N(\lambda-\frac{1}{2})\sigma_{\rm c}+O(\sigma_{\rm c}^2)\right)\ ,
\end{equation}
where we have kept only the first order correction term. 

If we assume that $\sigma_{\rm h}\ll 1$ and $a=\beta_{\rm h}/\beta_{\rm c}$ small, then the optimal value for the  frequency ratio is given by
\begin{equation}\label{Xopt}
 \overline{x}_{\text{sq}}=\left(\frac{a\beta_{\rm c} \langle H\rangle_A}{\N}\right)^{1/4}=\left(\frac{\beta_{\rm h}\langle H\rangle_A}{\N}\right)^{1/4}\ .
\end{equation}

Thus, one obtains the following expression for the optimal output work and efficiency
\begin{subequations}
\begin{eqnarray}
\tilde{W}_{\text{sq}}^{(\N,\lambda)} &=& \frac{\N}{2a\beta_{\rm c}}\left[\left(1-\sqrt{\alpha_{\rm sq}}\right)^2+\left(\frac{1-\sqrt{\alpha_{\rm sq}}}{2\alpha_{\rm sq}^{1/4}}\right)g_{\N}(\lambda-\frac{1}{2})a\sigma_{\rm c}+O(\sigma_{\rm c}^2)\right]\ ,\label{OptWorkGen} \\
\eta_{\text{sq}}^{(\N,\lambda)} &=& \frac{1-\sqrt{\alpha_{\text{sq}}}+\frac{a\sigma_{\rm c}}{2\alpha_{\text{sq}}^{1/4}}g_\N(\lambda-\frac{1}{2})}{2+\sqrt{\alpha_{\text{sq}}}+\frac{a\sigma_{\rm c}}{\alpha_{\text{sq}}^{1/4}(1-\sqrt{\alpha_{\text{sq}}})}g_\N(\lambda-\frac{1}{2})}\ ,\label{OptEffGen}
\end{eqnarray}
\end{subequations}
where we have introduced $\alpha_{\text{sq}}=\frac{a\beta_{\rm c} \langle H\rangle_A}{\N}$ and performed a very similar computation for the heat $\langle \mathrm{Q}_2\rangle$.

\begin{figure}[t]
\begin{center}
\includegraphics[scale=0.8]{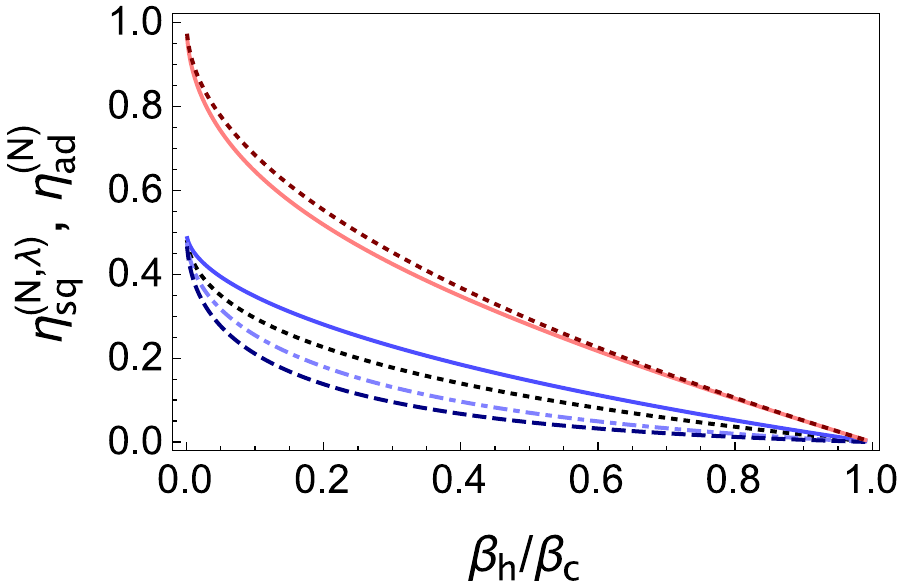}
\end{center}
\caption{\textbf{Sudden quench and adiabatic efficiency at optimal work of a many-particle engine}. The efficiency at optimal work as a function of $\beta_{\rm h}/\beta_{\rm c}$ for $\beta_{\rm c}=0.01/(\hbar \omega_1)$ is plotted. The two top curves  represent the adiabatic efficiency for $\N=1$ (dotted line) and for $\N=500$ (continuous line). The four curves below correspond to the efficiency for a sudden quench driving for $\N=1$ (dotted line) and for $\N=500$ (continuous line for $\lambda=0$, dotted-dashed line for $\lambda=1/2$ and dashed line for $\lambda=1$). We observe that the efficiency for an adiabatic driving is above the sudden quench efficiency as expected. The novelty concerns the variation of the efficiency with $\N$ and $\lambda$. It is clear that for $\lambda<1/2$ (between bosons and semions) the efficiency of the many-particle QHE is enhanced while it decreases for $\lambda\geq 1/2$ (semions, fermions and strongly correlated bosons). In the adiabatic case, the efficiency at optimal work does not depend on  the exclusion statistics (i.e., on $\lambda$) and decreases as a function of the particle number $\N$, see Fig. 2 in the main body of the manuscript.}
\label{figSM:figure3}
\end{figure}

\begin{figure}[t]
\begin{center}
\includegraphics[scale=0.8]{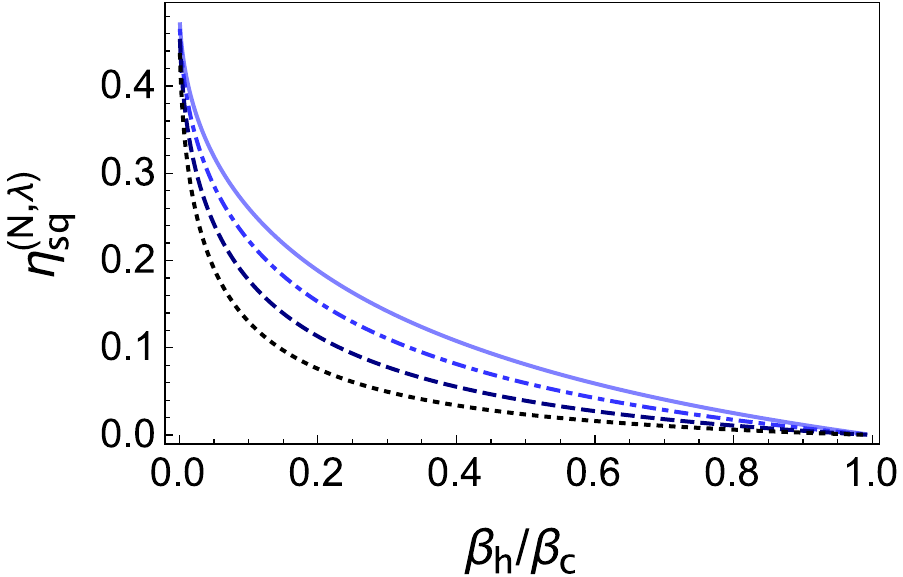}
\end{center}
\caption{\textbf{Efficiency at optimal work under sudden-quench driving for a many-particle QHE and a working medium  for interaction strength larger than $1$.} The efficiency is displayed for $\N=200,\ \beta_{\rm c}=0.01/(\hbar \omega_1)$, and for $\lambda=1$ (continuous line), $\lambda=2$ (dotted-dashed line), $\lambda=4$ (dashed line), $\lambda=8$ (dotted line). This figure illustrates the general result we proved concerning the fact that the efficiency is maximal for bosons and decreases for $\lambda>0$. For $\lambda>1$  the Calogero-Sutherland gas describes strongly-correlated bosons, similar to a super-Tonks-Girardeau gas \cite{Greg}. For large $\lambda$ the working medium becomes a Wigner crystal with spatially separated  particles. In this case it is clear that the many-particle QHE  is not efficient.}
\label{figSM:figure4}
\end{figure}

Explicit expressions  for the optimal work at different regimes of inverse temperature $\beta_{\rm c}$ can be found:  
\begin{itemize}
 \item For $\sigma_{\rm c}\gg \N$, one has $ \overline{x}_{sq}\approx \kappa_{\N,\lambda}\sqrt{\hbar\beta_{\rm h}\omega_1}$, where $\kappa_{\N,\lambda}=\sqrt{(1+(\N-1)\lambda)}$ for $\N\geq 1$. Then,
 \begin{subequations}
\begin{eqnarray}
\tilde{W}_{\text{sq}}^{(\N,\lambda)} &=& \frac{\N}{2a\beta_{\rm c}}\left(1-\kappa_{\N,\lambda}\sqrt{\hbar\beta_{\rm h}\omega_1/2}\right)^2\ , \label{OptWorkVLTN}\\
\eta_{\text{sq}}^{(\N,\lambda)} &=& \frac{1-\kappa_{\N,\lambda}\sqrt{\hbar\beta_{\rm h}\omega_1/2}}{2+\kappa_{\N,\lambda}\sqrt{\hbar\beta_{\rm h}\omega_1/2}}\ .\label{OptEffVLTN}
\end{eqnarray}
\end{subequations}
\item For large temperature $\sigma_{\rm c}\ll 1$,
\begin{itemize}
\item for $\N=1$, we have $\overline{x}_{\text{sq}}\approx a^{1/4}(1+\frac{1}{48}\sigma_{\rm c}^2)$, and
\begin{subequations}
\begin{eqnarray}
\tilde{W}_{\text{sq}}^{(1,\lambda)} &=& \frac{1}{\beta_{\rm c}}\left[\frac{(1-\sqrt{a})^2}{2a}+\frac{(-1+\sqrt{a}+a-a^{3/2})}{24\sqrt{a}}\sigma_{\rm c}^2\right]\ ,\label{OptWorkLT1}\\
\eta_{\text{sq}}^{(1,\lambda)} &=& \frac{1-\sqrt{a}}{2+\sqrt{a}}-\frac{(3-2a^{3/2})\sqrt{a}}{24(2+\sqrt{a})^2}\sigma_{\rm c}^2\ .\label{OptEffLT1}
\end{eqnarray}
\end{subequations}
\item for $\N>1$, $ \overline{x}_{\text{sq}}\approx a^{1/4}(1+\frac{1}{8}\sigma_{\rm c} g_\N(\lambda-\frac{1}{2}))+ O(\sigma_{\rm c}^2)$ and
\begin{subequations}
\begin{eqnarray}
\tilde{W}_{\text{sq}}^{(\N,\lambda)} &=& \frac{\N}{\beta_{\rm c}}\left[\frac{(1-\sqrt{a})^2}{2a}+\left(\frac{1-\sqrt{a}}{4\sqrt{a}}\right)(1-a^{1/4})(\frac{1}{2}-\lambda)g_\N\sigma_{\rm c} + O(\sigma_{\rm c}^2)\right]\ ,\label{OptWorkLTN}\\
\eta_{\text{sq}}^{(\N,\lambda)} &=& \frac{1-\sqrt{a}}{2+\sqrt{a}} + \frac{(3-2a^{3/4})\sqrt{a}}{4(2+\sqrt{a})^2}\left(\frac{1}{2}-\lambda\right)g_\N\sigma_{\rm c} + O(\sigma_{\rm c}^2)\ .\label{OptEffLTN}
\end{eqnarray}
\end{subequations}
\end{itemize}
\item For $\N>1$, we consider an intermediate regime corresponding to a large temperature $\hbar\beta_{\rm h}\omega_1\ll 1$ but a relatively small temperature per particle $\sigma_{\rm c}=\N\hbar\beta_{\rm c}\omega_1\geq 1$, which means that the number of particles is large $\sigma_{\rm h}\ll 1$ (i.e., $\beta_2\ll\beta_1$). Using \eqref{Int}, we find $\frac{a\beta_{\rm c} \langle H\rangle_A}{\N}\approx a\mu_\lambda(\sigma_{\rm c})$  
where $\mu_\lambda(\sigma_{\rm c})$ is given by \eqref{mu}. Notice that $\mu_0(\sigma_{\rm c})$ is a decreasing function of $\sigma_{\rm c}$ with the following bounds $\pi^2/(6N\hbar\omega_1\beta_{\rm c})\leq \mu(\sigma_{\rm c})\leq 1$. If one considers $\sigma_{\rm c}$ not too large (so that $\sigma_{\rm h}=\N\hbar\beta_{\rm h}\omega_2\ll 1$ still holds), by \eqref{Xopt} and \eqref{OptEffGen} one finds
\begin{subequations}
\begin{eqnarray}
\tilde{W}_{\text{sq}}^{(\N,\lambda)} &=& \frac{\N}{2a\beta_{\rm c}}\left(1-\sqrt{\mu_\lambda(\sigma_{\rm c}) a}\right)^2\ ,\label{OptWorkGenInt}\\
\eta_{\text{sq}}^{(\N,\lambda)} &=& \frac{1-\sqrt{\mu_\lambda(\sigma_{\rm c}) a}}{2+\sqrt{\mu_\lambda(\sigma_{\rm c}) a}}\ ,\label{OptEffGenInt}
\end{eqnarray}
\end{subequations}
where $\mu_\lambda(\sigma_{\rm c})$ is defined by \eqref{mu}. Thus, the optimal output power as well as the efficiency at optimal output power can be substantially improved for $\lambda<1/2$ (and for $\sigma_{\rm c}$ small enough) as $\mu_\lambda\leq 1$ and decreases otherwise (i.e., for $\lambda>1/2$), see Figs. \ref{figSM:figure1}, \ref{figSM:figure2}, \ref{figSM:figure3}, and \ref{figSM:figure4}. Notice that \eqref{OptWorkGenInt} and \eqref{OptEffGenInt} are valid only for $a\ll 1$ as $\sigma_{\rm h}\ll 1$. The  case $\sigma_{\rm h}\geq 1$ will be studied in the next section.
\end{itemize} 

\section{Many-particle QHE in the low-temperature limit of the hot reservoir}\label{Multipart}

So far we have exclusively considered the large temperature limit of the  hot reservoir, $\sigma_{\rm h}\ll1$, that imposes a restriction on the number of particles in the working medium. In this section we explore the properties of the QHE beyond this regime considering a larger number of particles $1\ll \sigma_{\rm c(h)}\ll \N$. Next, we show that for a sudden quench protocol the efficiency can be improved substantially for the weakly interacting case and decreases drastically otherwise. For an adiabatic driving we establish that the efficiency decreases with $\N$.  

\subsection{Sudden quench protocol}

\begin{figure}[t]
\begin{center}
\includegraphics[scale=0.8]{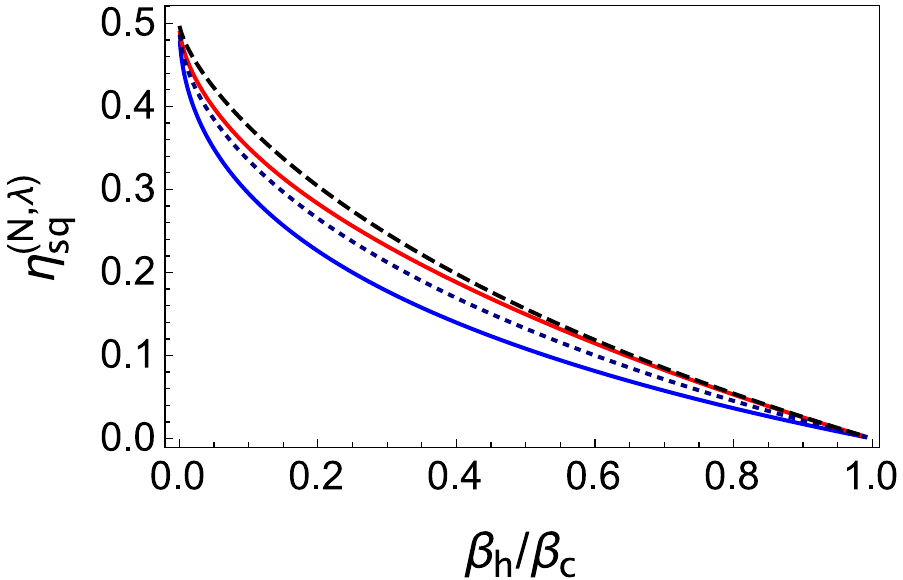}
\end{center}
\caption{\textbf{Efficiency at optimal work of a QHE with an ideal Bose gas as working medium.} In this figure we show the lower and upper bound for the efficiency at optimal work in the large temperature regime $1\leq \sigma_{\rm c(h)}\ll N$. The top continuous line is the numerical exact efficiency for $\N=1000$, and the bottom continuous line is for $\N=1$. The dotted line (respectively dashed line) is the lower bound (respectively upper bound) given by \eqref{EffLUB}. We remark that the efficiency can be improved drastically and a similar plot for the relative efficiency shows that it is enhanced between 25$\%$ (lower bound) and 50$\%$ (upper bound). We notice that for $\beta_{\rm h}/\beta_{\rm c}\gtrsim 0.4$ the upper bound is sharper than the lower bound and provides an accurate estimate of the efficiency. } 
\label{figSM:figure5}
\end{figure}

Numerically, we observe that for an ideal Bose gas (the optimal working medium) the efficiency grows up when $\N$ becomes large but saturates when $\N\hbar\beta_{\rm h}\omega_2 > 10$. To understand physically this curious phenomenon, it suffices to write the expression of the total work for large particle number $\N\gg (\hbar\beta_{\rm c(h)}\omega_{1(2)})^{-1}$ considering $\sigma_{\rm c(h)}\ll \N$ (i.e., $\hbar\beta_{\rm c}\omega_1\ll 1$) as we know that in the  very-low temperature regime of the cold reservoir (when $\hbar\beta_{\rm c}\omega_1\gg 1$) the engine does not operate efficiently. In this novel regime, we find that the total work does not depend on $\N$,
\begin{equation}\label{WorkNlarge}
\tilde{W}_{\text{sq}}=\frac{\pi^2}{12\hbar\beta_{\rm c}^2\omega_1}\left[\frac{x^2-1}{x^2}+(1-x^2)\frac{x}{a^2}\right]\ .
\end{equation} 
To derive this equation we use \eqref{Int} and the fact that $\mu_{0}(\sigma_{\rm c})\approx \frac{\pi^2}{6\sigma_{\rm c}}$ which leads to $\langle H\rangle_{A}\approx \frac{\N\pi^2}{6\beta_{\rm c}\sigma_{\rm c}}=\frac{\pi^2}{6\hbar\beta_{\rm c}^2\omega_1} $ and $\langle H\rangle_{C}\approx \frac{\N\pi^2}{6\N\hbar\beta_{\rm h}^2\omega_2}=\frac{\pi^2 x}{6a\hbar\beta_{\rm c}^2\omega_1}$ as $\beta_{\rm h}=a\beta_{\rm c}$ and $\omega_2=\omega_1/x$. We have neglected the ground energy contribution assuming that $\N\hbar^2\beta_{\rm c(h)}^2\omega_{1(2)}^2\ll 1$. In the case where the ground state energy dominates, it is clear that the medium is not efficient as it is effectively ``frozen'' (as previously discussed).  As a result, we consider a  large  particle number $\N$ compared to $(\hbar\beta_{\rm c}\omega_1)^{-1}$ but very small compared to $(\hbar\beta_{\rm c}\omega_1)^{-2}$. In Figure \ref{figSM:figure5} we take $\hbar\beta_{\rm c}\omega_1=0.01$ and $\N=1000$ that corresponds to this regime as  in this case $100\ll \N\ll 10000$.  
The work done $\tilde{W}_{\text{sq}}$ is of order 
$$\frac{\pi^2 \langle \text{n}\rangle}{12\beta_{\rm h}}\ ,$$
where $\langle \text{n}\rangle=(\hbar\beta_{\rm h}\omega_2)^{-1}$ is the average number of excited states per particle for the equilibrium state $C$. We point out that the number of excited states per particle remains small compared to the number of particle in this regime $\langle \text{n}\rangle \ll \N$  so that the total work is also small compared to the classical one as $\tilde{W}_{\text{sq}}\ll\frac{\N}{2\beta_{\rm h}}$.    

In this regime, one can find the frequency ratio for which the work is optimal by solving this polynomial equation
\begin{equation*}
p_a(x)\equiv 3x^5-x^3-2a^2,\ 0<a<1\ ,
\end{equation*}
where the solution only depends on the temperature ratio $a$. Notice that we expect $a$ to be greater than $0.1$ to keep $\N\gg (\hbar\beta_{\rm h}\omega_2)^{-1}$. The solution of this equation can not be expressed as a standard solution of a polynomial equation since the degree of the equation is greater than $4$. However, numerical analysis shows that an estimate of the real solution takes the form
$$\overline{x}_{\text{sq}}\approx a^{s},\ 1/4\leq s\leq 1/3\ ,$$
where $a^{1/4}$ (respectively $a^{1/3}$) gives the upper bound (respectively lower bound) of the root for $a>0.1$. 

Hence, after computing the heat $\langle \mathrm{Q}_{2}\rangle$ we obtain this expression for the efficiency, see Fig. \ref{figSM:figure5},
\begin{equation}\label{EffLUB}
\eta^{(\N,0)}_{\text{sq}}\approx (1-a^{2s})\frac{1-a^{2-3s}}{2-a^{2-3s}-a^{2-s}},\ 1/4\leq s\leq 1/3\ ,
\end{equation}
where the limit case $s=1/4$ (respectively $s=1/3$) gives a lower bound (respectively upper bound) to the numerical exact computations for $\N\geq 1000$. 
Numerically (see Fig. \ref{figure2}) we show that the efficiency can be  enhanced up to 50$\%$, see Fig. \ref{figSM:figure2}, and Fig. \ref{figSM:figure5}. 
We further notice that from \eqref{EffLUB}, and \eqref{OptEffLT1}, the efficiency in this regime is upper-bounded by $150\%$ of the value of the single-particle efficiency.

While for the ideal Bose gas the $\lambda$-term vanishes, for the general case the ground state contribution of $\lambda$ can reduce the enhancement of the efficiency. 
This observation follows from the strong dependence of the total work on $\lambda$,
\begin{subequations}\label{WorkNlargelambda}
\begin{eqnarray}
\tilde{W}_{\text{sq}}&\approx & \frac{x^2-1}{2x^2}\frac{\N}{\beta_{\rm c}}\mu_\lambda(\sigma_{\rm c})+\frac{1-x^2}{2}\frac{\N}{\beta_{\rm h}}\mu_\lambda(\sigma_{\rm h}),
\ \mathrm{for}\ 1\leq\sigma_{\rm c,h}\ll \N \label{WorkNlargelambda1}\\
&\approx & \frac{\pi^2\langle \text{n}\rangle}{12\beta_{\rm h}}\left[a^2\left(\frac{x^2-1}{x^2}\right)\left(1+\frac{3\lambda\sigma_{\rm c}^2}{\pi^2}\right)+x(1-x^2)\left(1+\frac{3a^2\lambda\sigma_{\rm c}^2}{\pi^2 x^2}\right)\right],\ \mathrm{for}\ 1 \ll\sigma_{\rm c,h}\ll \N\ , \label{WorkNlargelambda2}
\end{eqnarray}
\end{subequations} 
where  $\langle \text{n}\rangle=(\hbar\beta_{\rm h}\omega_2)^{-1}$. To derive equation \eqref{WorkNlargelambda2} we use \eqref{Int}.
In the regime of large number of particles ($\sigma_{\rm c(h)}\gg 1$), we estimate $\mu_\lambda$ as
$$\mu_\lambda(\sigma_{\rm c})\approx \frac{\pi^2}{6\sigma_{\rm c}}+\frac{\lambda}{2}\sigma_{\rm c}=\frac{\pi^2}{6\sigma_{\rm c}}\left(1+\frac{3\lambda\sigma_{\rm c}^2}{\pi^2}\right)\ .$$
Hence, one finds different regimes depending on the value of $\lambda$. For $\lambda\ll \lambda_{\rm c}=\pi^2/(3\sigma_{\rm c}^2)$ the thermal energy dominates and then the problem can be treated as the case  with $\lambda=0$. 
On the contrary, for $\lambda\gg \lambda_{\rm c}$ the thermal energy becomes negligible, thus the quantum fluid works less efficiently and becomes ``frozen'' for $\lambda=1/2$ and $1$, see Figs. \ref{figSM:figure2}, \ref{figSM:figure3} and  \ref{figSM:figure4}. 

To conclude, for a many-particle QHE in the low temperature regime nonadiabatic effects can enhance the efficiency over the single-particle value for $\lambda=0$ and for small $\lambda$ (typically $\lambda\ll \lambda_{\rm c}$). However, for $\lambda\geq 1/2$ the many-particle QHE is less efficient than the single-particle heat engine. In this case it is better to engineer a series of QHE with lower number of particles.

\subsection{Adiabatic driving}

Under adiabatic driving,  the total work is given by
\begin{equation}\label{WorkadNlarge}
\tilde{W}_{\text{ad}}=\frac{\pi^2}{6\hbar\beta_{\rm c}^2\omega_1}\left[\frac{x-1}{x}+(1-x)\frac{x}{a^2}\right]\ .
\end{equation} 
The derivation of equation \eqref{WorkadNlarge} is very similar to \eqref{WorkNlarge}. 

The frequency ratio for which the work is optimal is determined by solving this polynomial equation
\begin{equation*}
p_a(x)\equiv 2x^3-x^2-a^2,\ 0<a<1\ ,
\end{equation*}
whose  solution  depends only on the temperature ratio $a$. Here we take $a$ to be greater than $0.1$ so that $\N\gg (\hbar\beta_{\rm h}\omega_2)^{-1}$. One can find an analytic expression of the real solution of this polynomial equation and provide an estimate for $a\geq 0.1$,
$$\overline{x}_{\text{ad}}\approx 1 +\frac{a-1}{2}-\frac{(a-1)^2}{16}\ .$$
Inserting this value in the Otto efficiency we obtain
\begin{equation}\label{EffLUBad}
\eta^{(\N)}_{\text{ad}}\approx \frac{9-10a+a^2}{16}\ .
\end{equation}
Combining \eqref{OptWorkEffAdInt2} valid for small $a$ and \eqref{EffLUBad} we find that the efficiency is minimum for $a\sim 0.1$ and is equal to 70$\%$ of the single-particle efficiency which is consistent with the numerical exact calculation  shown in  Fig. \ref{figSM:figure2}.


\end{document}